\documentclass[preprint,aip,jcp,superscriptaddress]{revtex4-1}

\usepackage{graphicx,color}
\usepackage{amsmath,amsfonts,amssymb}
\usepackage{dcolumn}

\newcommand{\fig}{Fig.}

\newcommand{\figref}[1]{\fig~\ref{#1}}

\newcommand{\tabref}[1]{table~\ref{#1}}

\renewcommand{\eqref}[1]{equation~(\ref{#1})}

\newcommand{\new}[1]{{#1}}

\newcommand{\vdw}{[H$_2 \cdots$OH] }

\begin{document}


\title{Reaction Rates and Kinetic Isotope Effects of
H$_2$ + OH $\rightarrow$ H$_2$O + H}


\author{Jan Meisner}
\affiliation{Institute for Theoretical Chemistry, University of Stuttgart, Pfaffenwaldring 55, 70569 Stuttgart,Germany}

\author{Johannes K\"{a}stner}
\affiliation{Institute for Theoretical Chemistry, University of Stuttgart, Pfaffenwaldring 55, 70569 Stuttgart,Germany}


\date{\today}

\begin{abstract}
We calculated reaction rate constants including atom tunneling of the reaction
of dihydrogen with the hydroxy radical down to a temperature of
50~K. Instanton theory and canonical variational theory with 
microcanonical optimized multidimensional tunneling (CVT/$\mu$OMT)
were applied using a fitted potential energy surface 
[J. Chem. Phys. 138, 154301 (2013)].
All possible
protium/deuterium isotopologues were considered.
\new{Atom tunneling increases at about 250~K (200~K for deuterium transfer)}.
Even at 50~K the rate constants of all isotopologues remain in the
interval $4\cdot 10^{-20}$ to $4\cdot 10^{-17}$ cm$^3$~s$^{-1}$,
demonstrating that even deuterated versions of the title reaction are possibly
relevant to astrochemical processes in molecular clouds. 
The transferred hydrogen atom dominates the kinetic isotope effect at all temperatures.
\end{abstract}


\maketitle

\section{Introduction}

The reaction of H$_2$ + OH has emerged as a prototype reaction for four-atomic
systems.  It contributes to fundamental processes in atmospheric chemistry,
astrochemistry, and combustion.\cite{pie09, rah03, gli15} The reaction being
surface catalyzed was shown to be one of the main routes of H$_2$O formation
in the interstellar medium.\cite{cup07,cup10,lam13a,lam14}

On surfaces the reaction was observed at the cryogenic temperature of 10~K
through quantum mechanical tunneling of atoms.\cite{oba12,mei16} 
\new{For the gas phase reaction, a} number of studies on this reaction, 
theoretical\cite{bha10, bha11, esp10, ngu11, fu10, fu15, cha04, man00} as well as experimental
\cite{rav81,tal96,kra04,ork06,lam13} down to 200~K has been performed. For an
overview of previous experimental and theoretical results, we refer to
reviews.\cite{cas02,smi02a}

In this article we present reaction rate constants of the title reaction down
to 100~K using instanton theory\cite{lan67,mil75,col77,cal77,gil77} and down
to 50~K using CVT/$\mu$OMT.\cite{liu93a,Truhlar_Faraday1994} Instanton
theory\new{\cite{aff81,col88,han90,
    ben94,mes95,ric09,alt11,rom11,rom11b,ric16}} is a semiclassical theory
based on Feynman's path integrals.\cite{fey48} It takes multidimensional
tunneling into account while only the optimization of a tunneling path -- the
instanton -- is necessary.\cite{kae14} Instanton theory is meanwhile
frequently used to calculate reaction rates in different areas of
chemistry.\cite{cha75,mil94,mil95,mil97,sie99, sme03,qia07,and09,
  gou10a,gou11,gou11b, rom11,gou10,jon10,mei11,gou11a,ein11,rom12,
  kry12,kae13,alv14,kry14} Canonical variational theory \new{(CVT) minimizes
recrossing compared to transition state theory (TST). It} was used with
microcanonical optimized multidimensional tunneling
(CVT/$\mu$OMT)\cite{liu93a,Truhlar_Faraday1994} was used along with zero
curvature tunneling (ZCT),\cite{gar80,tru83} small curvature tunneling
(SCT),\cite{sko81} large curvature tunneling (LCT),\cite{gar83,gar85,fer01}
and microcanonical optimized multidimensional tunneling
($\mu$OMT)\cite{liu93a,Truhlar_Faraday1994} calculations down to 50~K.  ZCT
assumes no deviation of the tunneling path from the classical minimum energy
path.  Compared to that, SCT considers corner-cutting effects and LCT
approximates the tunneling path by a linear path from reactants' valley to
products' valley. The $\mu$OMT method takes into account that the tunneling
path depends on the energy by using the maximum of SCT and LCT tunneling
probabilities at each energy.\cite{fer06,fer07}

\begin{figure}[h!]
\begin{center}
  \includegraphics[width=8cm]{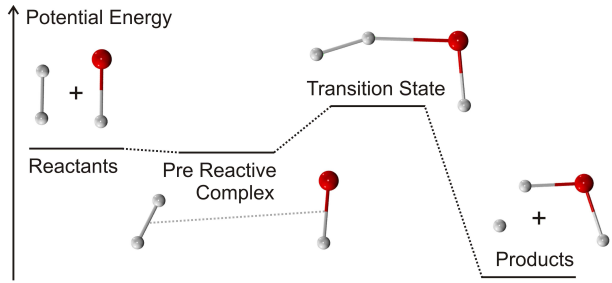}
  \caption{Potential energy profile of the reaction H$_2$ + OH $\rightarrow$
    H$_2$O + H. Relative to the separated reactants, the pre-reactive complex
    has a potential energy of $-$2.1~kJ~mol$^{-1}$, the transition state
    22.5~kJ~mol$^{-1}$ and the separated products $-$68.1~kJ~mol$^{-1}$.
    \label{reactionprofile}
  }
\end{center}
\end{figure}

As the four-atomic system of H$_2$ + OH is of fundamental interest, a variety
of potential energy surfaces (PES) have been published.\cite{och98, wu00,
  wal80,sch80,yan01,bet00} Recently, a global potential energy surface fitted
by a neural network to UCCSD(T)-F12a/AVTZ data was published (NN1
PES).\cite{che13} This PES was shown to give reliable results in, e.g.,
the study of the mode specificity of the H + HOD reaction. \cite{fu15} The NN1
PES was therefore applied here as well.  Although several studies on thermal
rate constants for the title reaction appeared, \cite{ngu11,man00,mat98} for
instance, the semiclassical transition state theory (SCTST) calculations of
Nguyen et al. \cite{ngu10} -- even to investigate reaction rate constants of
all isotopologues \cite{ngu11} -- it seems that this is the first study which
provides rate calculations on the NN1 PES.
\new{The reaction profile consisting of the stationary points are shown in
\figref{reactionprofile}.}

For bimolecular reactions, it is in general possible that a (weakly) bound
van-der-Waals complex can lead to an increase of the bimolecular reaction rate
constant with decreasing temperature.  This effect was studied experimentally
in the reaction of HBr + OH as well as in the reactions with nitric acid or
alcohols and OH radicals. \cite{ree15, bro01, sha13} In these cases, the
non-covalent interactions between the two reactants stem from the dipole
moments and polarizabilities of the reacting molecules.  
\new{In contrast to these,
H$_2$ is less polarizable and has no permanent dipole moment.
Thus, the intermolecular interaction between H$_2$ and the OH radical 
and the impact of the pre-reactive complex (PRC) \vdw are
expected to be small unless the temperature is much lower than
considered in this work.}

In this study we investigate the temperature dependence of the reaction rate
constant and compare it to published values.  Furthermore, the temperature
dependence for the rate constants for all eight possible isotopologue
reactions and the resulting kinetic isotope effects (KIEs) have been studied.
\new{At low temperatures (below $T_\text{c}$) tunneling dominates} the reaction
rate. The nuclear mass has a high impact on the tunneling probability leading
to large kinetic isotope effects (KIEs).

\section{Methods}

In instanton theory, the instanton, the saddle point corresponding to the
transition state, is a closed Feynman path folded back onto itself which spans
the barrier region. At high temperature it is short and covers only the top of
the barrier while at low temperature it protrudes right into the reactant
state region. At the crossover temperature,
\begin{equation}
 T_\text{c}  = \frac{\hbar \omega_\text{TS}}{2 \pi k_\text{B}},
 \label{eq:2}
\end{equation}
the instanton path generally collapses to a point and the theory becomes
inapplicable although extensions above the crossover temperature exist.\cite{zha14}
Here $\omega_\text{TS}$ is the absolute value of the imaginary
frequency at the transition state, $\hbar$ is the reduced Planck's constant,
and $k_\text{B}$ is Boltzmann's constant. $T_\text{c}$ is mass-dependent: for
the title reaction containing protium only is was found to be 276.2~K, while
204.2~K was found for the per-deuterated reaction. In many cases
$T_\text{c}$ can be used as a cheap and simple indication if atom tunneling is
important at the temperature of interest. Following \eqref{eq:2}, whenever
$\omega_\text{TS}$ is larger than 1300~cm$^{-1}$ atom tunneling is relevant at
room temperature.

The different H/D isotopologues are labeled as H$^1$H$^2$OH$^3$ such that the
reaction reads $\textnormal{H}^1\textnormal{H}^2+
\textnormal{O}\textnormal{H}^3 \rightarrow \textnormal{H}^1 +
\textnormal{H}^2\textnormal{O}\textnormal{H}^3$. DDOH therefore corresponds
to a reaction of OH with D$_2$ while HDOH corresponds to the reaction HD + OH
$\rightarrow $ H + DOH.

Vibrational modes were described by the harmonic approximation of the Feynman
path. The translational partition function was in all cases approximated by
the one of the ideal gas, which is identical to that of a quantum mechanical
particle in a box. The rotational partition function of the transition state
was obtained as the geometric mean value of the rotational partition functions
of all images along the instanton path treated as rigid quantum rotors. The
reactant molecules were, equivalently, treated as rigid rotors. The symmetry
number, the order of the rotational subgroup in the molecular point
group,\cite{fer07a} of the individual molecules was taken into account in the
rotational partition function, i.e. the one of H$_2$ and D$_2$ was divided by
two, while the one for HD is not.

The kinetic isotope effects are dominated by tunneling and by the zero-point
vibration. Neither the rotational nor the translational contribution have a
significant effect on the KIEs.

The NN1 PES\cite{che13} was interfaced with DL-FIND.\cite{kae09a}
Instantons were optimized starting from the classical transition state 
or by starting from an already optimized instanton of similar temperature using the adapted 
Newton--Raphson algorithm implemented in DL-FIND. \cite{rom11,rom11b}
The convergence criteria for the instanton optimization on the NN1 PES was $5\cdot 10^{-11}$
atomic units for the maximal component of the gradient. Note, that we use mass-weighted
coordinates and gradients with the masses in atomic units, i.e. relative to
the electron mass. This influences the convergence criterion. 

The instanton is a closed Feynman path with images having pairwise identical
coordinates.  The full path was represented by 512~images. Convergence with
respect to the number of images was tested at the most severe case with the
largest distances between adjacent images, the all-H reaction (HHOH) at 100~K.
In this case, the rate constant obtained with 4096 images for the full path
deviated by only 0.4~\% from the value obtained with 512~images \new{and is
  only 2.4~\% higher compared to the value obtained with 194~images.} Smaller
deviations can be expected at higher temperature or for heavier
isotopologues. Thus, we consider the discretization to be converged with
respect to the number of images.

To test the quality of the NN1 PES we additionally calculated instanton rate
constants with on-the-fly energy calculations at the CCSD(T)-F12
level\cite{adl07,adl09} using the cc-pVDZ-F12 basis set.\cite{pet08} The
program package Molpro version 2012\cite{MOLPRO2012,wer11} interfaced to
DL-FIND\cite{kae09a} via ChemShell\cite{met14} was used for these
calculations. Due to the high computational demands of these calculations, 194
images were used and the
instanton optimizations were considered converged for all absolute gradient
components smaller than $10^{-8}$~a.u.

\begin{figure}[h!]
  \begin{center}
    \includegraphics[width=16cm]{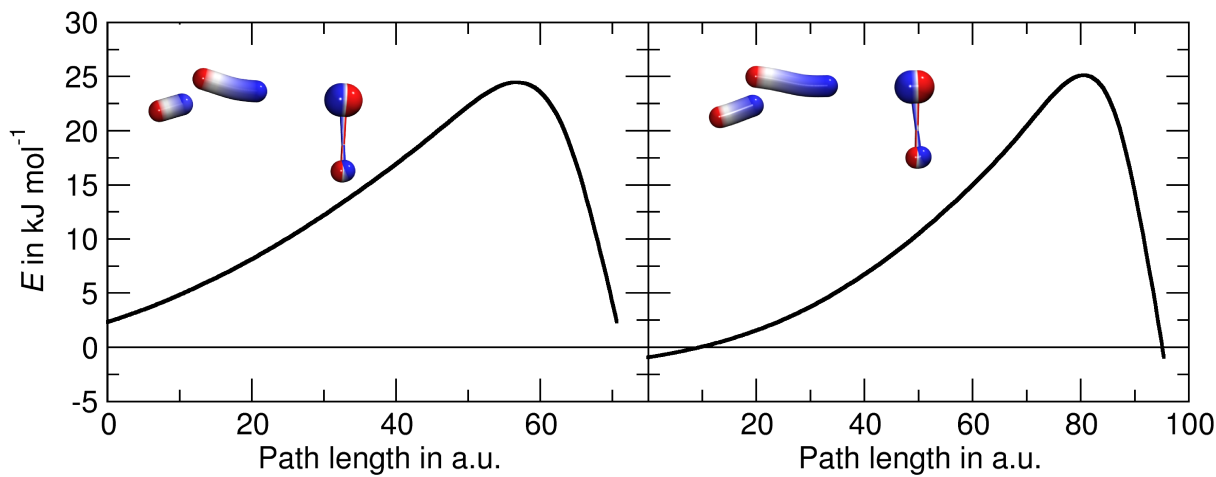}
    \caption{Potential energy along the instanton path at
      130~K (left) and 80~K (right) relative to the energy of the
      separated reactants. At 130~K the whole instanton path is above
      the reactant's energy, at 80~K its ends are below that value.
      Images of the corresponding instantons are inserted.
      \label{fig:Instanton}
    }
  \end{center}
\end{figure}

Below 100~K, the instanton path for HHOH stretches into the pre-reactive minimum with
parts of the path below the energy of the separated reactants, see
\figref{fig:Instanton}. 
\new{Instanton rates are not valid for energies below the separated reactants, so 
instanton rates are reported only down to 100~K for H-transfer.}
For the D-transfer the
whole instanton path remains above the reactants' energy for $T \textgreater
80$~K. Thus, instanton rates for D-transfer reactions are reported down to
80~K.
\new{CVT/$\mu$OMT was used down to 50~K.}

The ZCT, SCT, LCT, and $\mu$OMT calculations on the NN1 PES\cite{che13} have
been performed using POLYRATE \new{2010}\cite{lu92,POLYRATE2010} based on canonical
variational transition state theory (CVT).\cite{fer06,fer07} For the LCT
calculations, the action integrals ($\theta$ integrals) and the sine of the
angle between the minimum energy path and the tunneling path were interpolated
to 2$^{nd}$ order.

\section{Results}

\subsection{Reaction Rate Constants}

The relevant stationary points on the potential energy surface of the title
reaction are depicted in \figref{reactionprofile}. Relative to the separated
reactants H$_2$ and OH, the potential energy on the NN1 PES is
$-2.11$~kJ~mol$^{-1}$ for the PRC, $22.50$~kJ~mol$^{-1}$ for the transition
state (TS) and $-68.08$~kJ~mol$^{-1}$ for the products (H + H$_2$O).
The relative energies of the corresponding stationary points 
optimized on CCSD(T)-F12/cc-pVDZ-F12 level
are $-$1.77 kJ/mol for the PRC, 23.98 kJ/mol for the TS and $-$64.94 kJ/mol for 
the products.
The imaginary harmonic frequency is 1206~i~cm$^{-1}$ on the NN1 PES and 
1199~i~cm$^{-1}$ on CCSD(T)-F12/cc-pVDZ-F12 level.

\begin{figure}[h!]
  \begin{center}
    \includegraphics[width=8cm]{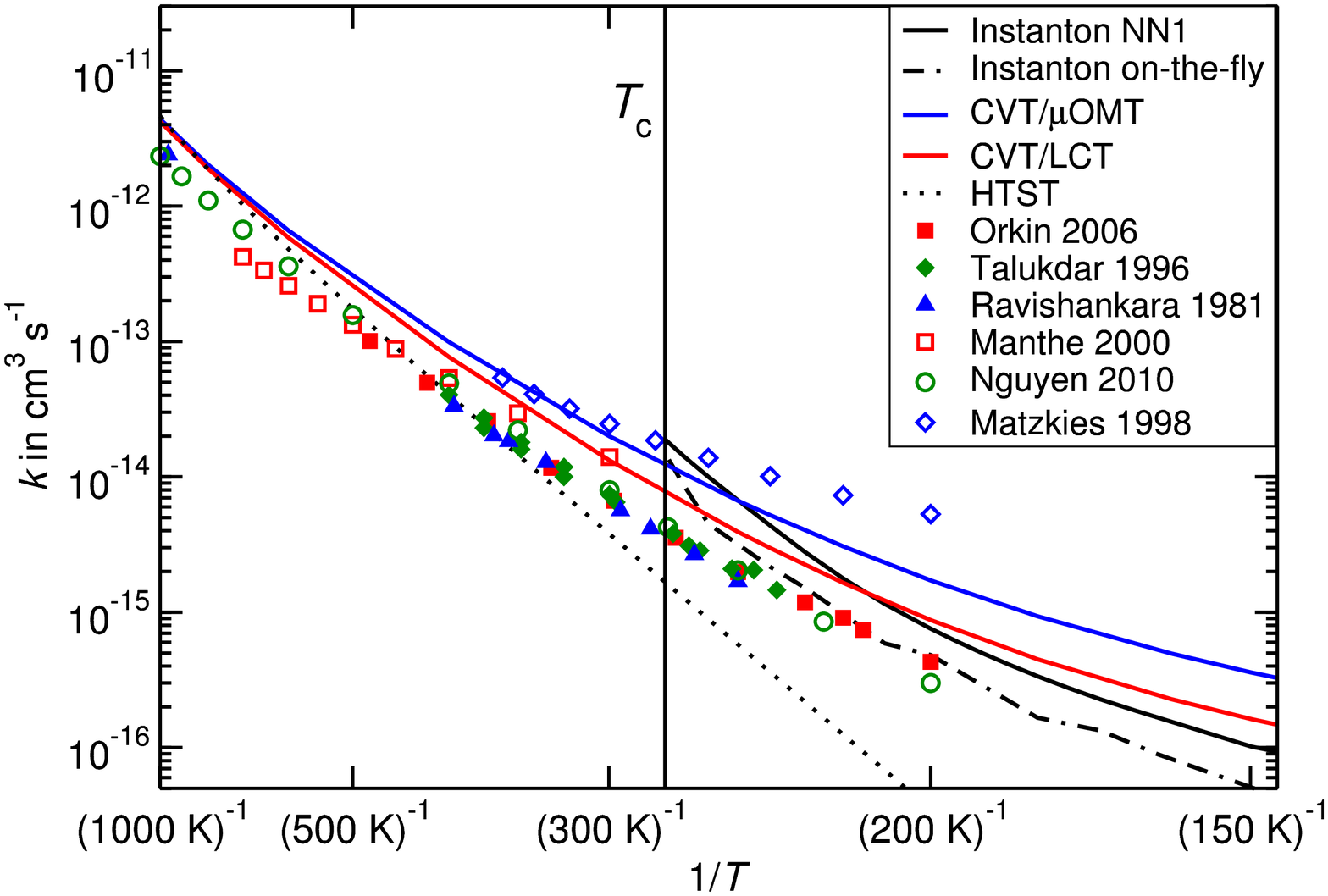}
    \caption{Reaction rate constants for HHOH 
      compared to literature data. Experimental data: 
      ``{\textcolor{red}{$\blacksquare$}}'' data from \cite{ork06},
      ``{\textcolor{green}{$\blacklozenge$}}'' data from \cite{tal96},
      ``{\textcolor{blue}{$\blacktriangle$}}'' data from \cite{rav81};
      computational data:
      ``{\textcolor{red}{$\square$}}'' data from \cite{man00},
      ``{\textcolor{green}{$\bigcirc$}}'' data from \cite{ngu10},
      ``{\textcolor{blue}{$\Diamond$}}'' data from \cite{mat98}.
      \label{fig:rates}
    }
  \end{center}
\end{figure}

The rate constant of the title reaction has been measured several times using 
different techniques, see \figref{fig:rates}. An Arrhenius plot shows a
noticeable curvature already at 300~K and below,\cite{ork06,rav81} which is a
clear sign that the reaction is influenced by atom tunneling. Experimental
rate constants are available from 1000~K down to
200~K.\cite{tal96,ork06,lam13,rav81,kra04} The different sets agree quite
well, typically within 20--30\% of the rate constant.

Among the computational studies of this system, Matzkies and Manthe
\cite{mat98,man00} performed full-dimensional quantum dynamics calculations on
the Schatz--Elgersma PES \cite{wal80,sch80} using the multi-configuration
time-dependent Hartree \new{(MCTDH)} approach. At $T=300$ K, the lowest temperature
covered by close-coupling calculations employing a rigorously correct
statistical sampling scheme for the rotational degrees of freedom,\cite{man00}
their calculations overestimate the experimental rate constants by about a
factor of 2. In an earlier work,\cite{mat98} they calculated rate constants
down to 200 K, which are more than an order of magnitude higher than the
experimental value.\cite{ork06} Better agreement with the experimental values
was achieved by Nguyen et al.\cite{ngu10} by applying semiclassical
transition-state theory (SCTST) on high-level direct-dynamics energies. At
200~K they underestimate the experimental rate constant by a factor of 1.43,
which is comparable to the experimental error bar. They furthermore showed that
SCT gives significantly higher reaction rate constants at lower temperatures
compared to SCTST calculations. \cite{ngu10}

For the title reaction containing protium only (HHOH), the crossover
temperature is $T_\text{c}=276.2$ K. Using instanton theory, we calculated
bimolecular reaction rates using the NN1 PES at $T=270$ K and below as
it is only applicable below $T_\text{c}$.  The results are
depicted in \figref{fig:rates}, numbers are given in the supporting
information.\cite{ThisSI} As expected,\cite{gou11a} instanton theory overestimates the
rate constant close to $T_\text{c}$. Agreement is improved at lower temperature. At
220~K our rate constant ($1.78\cdot10^{-16}$~cm$^3$~s$^{-1}$) is higher by a
factor of 1.95 than the results of flash photolysis resonance-fluorescence by
Orkin et al.;\cite{ork06} at 200~K, the lowest temperature at which comparison is
possible, it is still higher by a factor of 1.76. 
A higher accuracy of
instanton theory can be expected at lower temperature due to the known
overestimation close to $T_\text{c}$.\cite{gou11a}

The deviation might reflect a deficiency of the rate theory or the
potential. To test the accuracy of the potential, we recalculated instantons and
rate constants on-the-fly on CCSD(T)-F12/cc-pVDZ-F12 level without fitting the
PES.  The reaction rate constants obtained in this way agree better with the
experimental result, 
overestimating it by a factor of 1.49 at 240~K and only by a factor of 1.12 at 200~K.
\new{
Thus, the NN1 PES leads to a slight overestimation of the rate constants.
However, we continue with the NN1 PES as 
on-the-fly calculations for all isotopologues would be too costly.
}
While the absolute rate constants might be overestimated by a factor of 
approximately 1.5 to 2.0, one can assume the KIEs incur smaller errors. We
assume roughly the same level of accuracy at low temperatures, where no other
experimental or computational data are available for comparison.

Given that instanton theory is expected to
be more accurate at lower temperatures, the rates using the NN1 PES, which
employs a better basis set, are more promising than the direct-dynamics
results. Still, the NN1 PES probably shows slight inaccuracies in the region
of the configurational space most relevant to tunneling, leading to a slight
overestimation of the reaction rate. However, we continue with the NN1
PES. While the absolute rate constants might be overestimated by a factor of 
approximately 1.5 to 2.0, one can assume the KIEs incur smaller errors. We
assume roughly the same level of accuracy at low temperatures, where no other
experimental or computational data are available for comparison.

To test the accuracy of the rate theory, we performed CVT/ZCT, SCT, LCT,
and $\mu$OMT calculations on the NN1-PES.  As in
this reaction the SCT rate constant is always higher than the one
obtained by LCT, the $\mu$OMT result is virtually indistinguishable from the results
obtained by SCT. Therefore no graph for SCT is shown in
\figref{fig:rates}.  At temperatures below 300~K, CVT/LCT (and ZCT)
agrees well with instanton theory whereas SCT, and thus $\mu$OMT, give
significantly higher rate constants.  At 200~K, $\mu$OMT overestimates
the reaction rate constants by a factor of 4.0, see
\tabref{tab:rates} and \figref{fig:rates}. 

For comparison, \figref{fig:rates} includes the rate constant calculated
\new{by TST with all vibrations treated via quantum partition functions of
  harmonic oscillators, i.e., accounting for the vibrational zero-point energy
  but not for tunneling (Harmonic transition state theory, HTST)}.  As
expected, it describes the rate constant very well at high temperatures (close
to and above 400 K) but deviates significantly below about 300 K.

\begin{table}[ht!]
 \caption{
 \label{tab:rates}
   Reaction rate constants $k$ in cm$^3$~molecule$^{-1}$~s$^{-1}$ at 200~K obtained by different methods. 
   Experimental values are from reference \citenum{ork06}.
   }
    \begin{center}
    \setlength{\tabcolsep}{1mm}
\begin{tabular}{ll}
\hline
Exp.				&	4.3 $\cdot 10^{-16}$	\\
Instanton NN1			&	7.56$\cdot 10^{-16}$	\\
Instanton on-the-fly		&	4.83$\cdot 10^{-16}$	\\
CVT/$\mu$OMT			&	1.73$\cdot 10^{-15}$	\\
CVT/SCT				&	1.73$\cdot 10^{-15}$	\\
CVT/LCT				&	8.77$\cdot 10^{-16}$	\\
CVT/ZCT				&	7.00$\cdot 10^{-16}$	\\
CVT				&	2.76$\cdot 10^{-17}$	\\
\new{H}TST			&	3.25$\cdot 10^{-17}$	\\
\hline
\end{tabular}
\end{center}
\end{table}

\subsection{Kinetic Isotope Effects}

\begin{figure}[h!]
  \begin{center}
    \includegraphics[width=8cm]{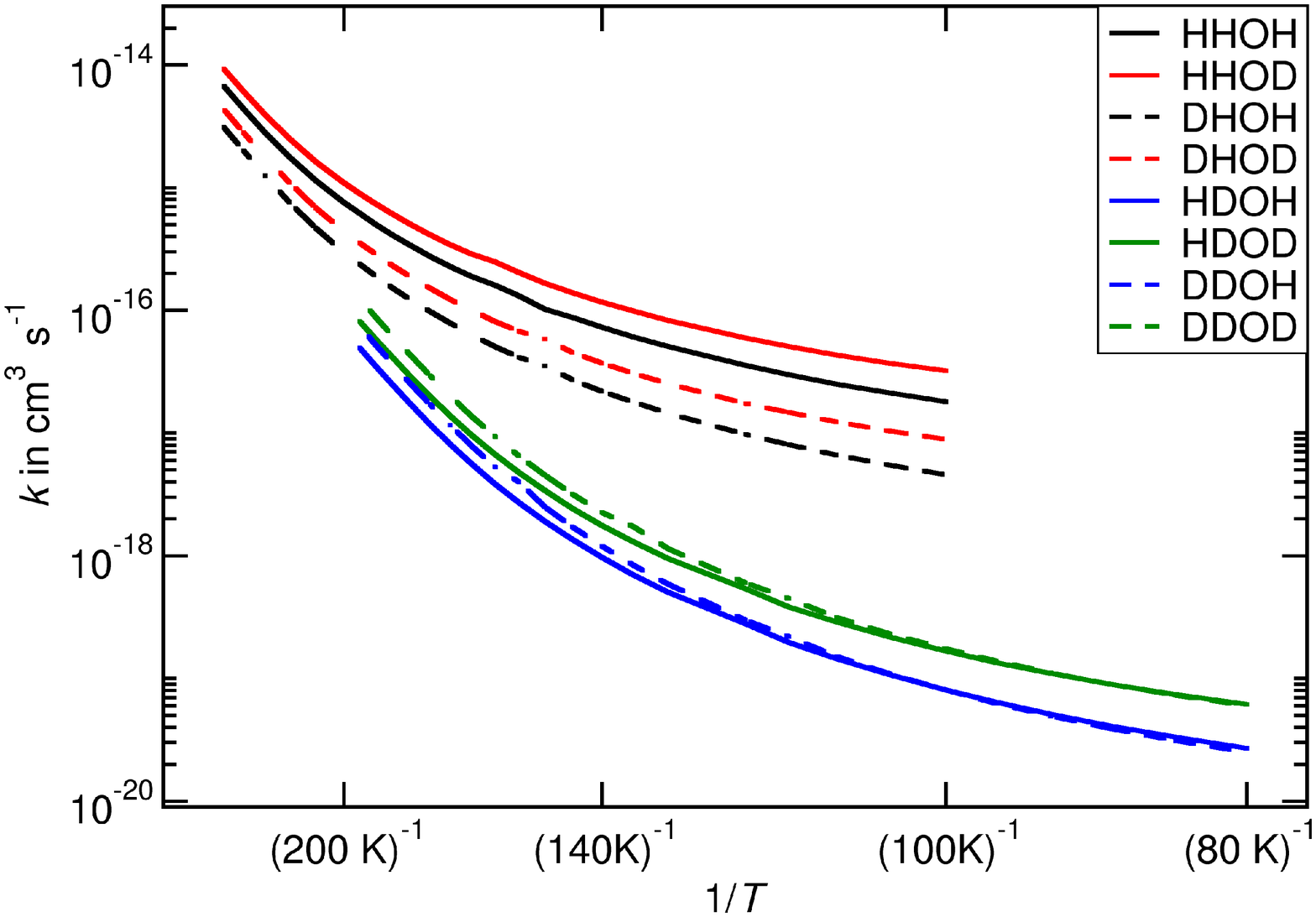}
    \includegraphics[width=8cm]{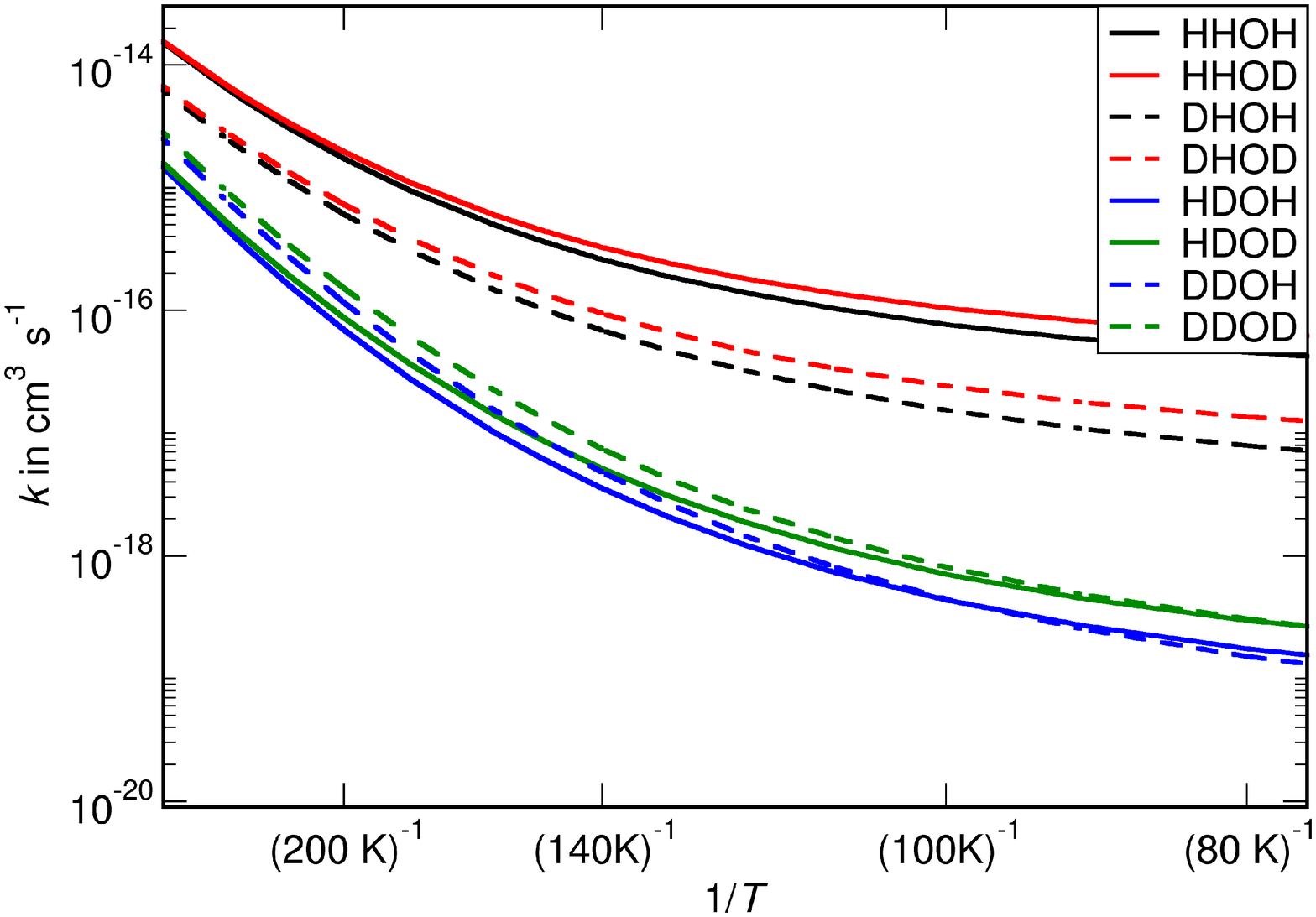}
    \caption{Temperature dependence of the reaction rate constants of
      all H/D isotopologues calculated with the instanton method (left) and
      with CVT/$\mu$OMT (right). 
      \label{fig:iso}
      }
  \end{center}
\end{figure}

All eight possible isotopologues were investigated. The zero-point
energy (ZPE) corrected energies of PRC and TS, as well as $T_\text{c}$
are given in \tabref{tab:ZPE}, the rate constants are shown in
\figref{fig:iso}. Values of the KIEs at 160~K and 100~K (both instanton
and $\mu$OMT), and 50~K ($\mu$OMT) are given in \tabref{tab:KIE}.

For reactions with deuterium, the crossover temperature
is significantly reduced, see \tabref{tab:ZPE}.
Down to 120~K the curvature of the
resulting Arrhenius plot in \figref{fig:iso} is negligible for isotopologues
with D-transfer.  Defazio et al. already mentioned that tunneling may not be
very important in the DDOH case.\cite{def03} This is certainly true in the
temperature range where experimental data is available, i.e. above 210~K.
\new{At lower temperature the reactions of all isotopologues are dominated by tunneling.}
A direct comparison between our instanton calculations and experimental data is
impossible for any of the deuterated cases, as no data is available below
$T_\text{c}$. Above 50 K a clear primary KIE, i.e., depending on the mass of
the atom to be transferred, is measurable. 


\begin{table}[h!]
 \caption{
 \label{tab:ZPE}
   ZPE corrected energies of the corresponding characteristic points of the
   PES in~kJ~mol$^{-1}$ relative to the separated reactants. The crossover
   temperature $T_\text{c}$ is given in K. $E_\text{a}$ refers to the
   activation energy, the energy difference between TS and PRC.}
    \begin{center}
    \setlength{\tabcolsep}{1mm}
\begin{tabular}{crccccr}
\hline
	&	PRC	&	TS	&	$E_\text{a}$	&
$E_\text{a}$(ref \cite{ngu11})&	$T_c$		\\
\hline                                                                                         
HHOH	&	0.54	&	24.76	&	24.22	&	24.41	       &	276.2	\\
HHOD	&	0.41	&	23.50	&	23.09	&	23.19	       &	276.1	\\
DHOH	&	0.30	&	24.13	&	23.82	&	23.74	       &	266.0	\\
DHOD	&	0.17	&	22.84	&	22.67	&	22.48	       &	265.8	\\
HDOH	&	0.20	&	25.86	&	25.66	&	25.37	       &	208.9	\\
HDOD	&	0.04	&	24.57	&	24.53	&	24.12	       &	208.8	\\
DDOH	&	$-$0.04	&	25.65	&	25.69	&	25.16	       &	204.3	\\
DDOD	&	$-$0.22	&	24.31	&	24.53	&	23.86	       &	204.2	\\
\hline
\end{tabular}
\end{center}
\end{table}

The KIE can stem from differences in zero-point energies or from
tunneling.\cite{per12a,sul13, per14} One may, of course, argue if the
harmonic approximation for zero-point energies is good enough to
estimate rate constants at such low temperatures. However, our
calculated vibrationally adiabatic barriers of the isotopologues agree
well (deviation $<0.7$~kJ~mol$^{-1}$) from the literature values
obtained by the more elaborate HEAT protocol,\cite{ngu11} see
\tabref{tab:ZPE}. It was shown previously \cite{ngu11} that including
anharmonicity changes the corresponding barrier height by less than
0.33~kJ~mol$^{-1}$.

Apart from the primary KIE, we observe that deuteration of the hydroxy
radical (OD) increases the reaction rate, leading to inverse KIEs.
Depending on the deuteration of the other sites, OD increases the
rates by factors of 1--3, see \tabref{tab:KIE}.  The main reason for
this effect is that the heavier deuterium atom lowers $E_\text{va}$ of
the transition state by reducing the zero-point energy of the
deformation modes of the two molecules with respect to each other.

\begin{table}[ht!]
  \caption{
    \label{tab:KIE}
    Kinetic isotope effects at \new{160~K}, 100~K, and 50~K with respect to HHOH.
  }	
  \begin{center}
    \begin{tabular}{l| D{.}{.}{3}D{.}{.}{3}| c D{.}{.}{3}D{.}{.}{3}D{.}{.}{3}}
  \hline
        	&\multicolumn{2}{c|}{Instanton} &&\multicolumn{3}{c}{$\mu$OMT}\\
  Isotopes	&\multicolumn{1}{c}{160 K} &\multicolumn{1}{c|}{100 K} 
		&\,&\multicolumn{1}{c}{160 K} &\multicolumn{1}{c}{100 K} &\multicolumn{1}{c}{50 K} \\
  \hline
HHOH	&	1.00	&	1.00	&  &     1.00         &  1.00         &  1.00  \\
DHOH	&	3.19	&	3.95	&  &     3.37         &  4.99         &  6.89  \\
HHOD	&	0.649	&	0.561	&  &     0.828        &  0.732        &  0.610  \\
DHOD	&	1.99	&	2.04	&  &     2.58         &  3.19         &  3.42  \\
HDOH	&	41.5	&	224	&  &     49.1         &  176          &  382  \\
DDOH	&	30.2	&	225	&  &     32.4         &  172          &  558  \\
HDOD	&	23.9	&	108	&  &     35.4         &  109          &  216  \\
DDOD	&	17.1	&	104	&  &     22.3         &  95.3         &  229  \\
\hline  
    \end{tabular}
  \end{center}
\end{table}

The reaction rate constants obtained with $\mu$OMT are higher than the ones
obtained with instanton theory by a factor of 4.2 for HHOH and 5.4 for HDOH at
100~K. It is obvious from \figref{fig:rates} and \tabref{tab:rates} that
CVT/$\mu$OMT generally overestimates the reaction rate constants for this
reaction, see also Fig. S1\new{, because in $\mu$OMT the tunneling path is not
optimized}.  Apart from that, the rate constants seem to follow the same
trends, in particular the KIEs obtained by both methods agree reasonably well,
see \tabref{tab:KIE}.

Instanton theory provides a dominant tunneling path for each specific
temperature. At low temperature, that path is almost
temperature-independent. The atoms contribute quite differently to that
tunneling path. Geometries and the energy along the instanton path are depicted in
\figref{fig:Instanton}. In the low-temperature limit for HHOH, the hydrogen
atom to be transferred is delocalized over 1.34~\AA{}, the one that remains as
isolated hydrogen atom after the reaction over 0.80~\AA{}. Both oxygen and
hydrogen of OH contribute to the tunneling much weaker, they are delocalized
over 0.14 and 0.21~\AA{}, respectively. Deuteration changes these
contributions: for HDOH, the transferred deuterium is delocalized over only
1.25~\AA{} while the other tunneling path length remain almost unchanged
(0.77, 0.15, and 0.21~\AA).

We found primary H/D-KIEs of $>200$ at 50~K using CVT/$\mu$OMT. At even lower
temperature than reported here, the KIE can be expected to be at least
similarly strong. Consequently we expect a significant influence of this
reaction and its KIE on the deuterium fractionation of molecules in the
interstellar medium.

\section{Summary}

We calculated reaction rate constants of H$_2$ + OH $\rightarrow$ H + H$_2$O
down to 100~K using instanton theory and down to 50~K using CVT/$\mu$OMT on
the NN1 PES \cite{che13} for all H/D isotopologues. Atom tunneling sets in at
about 250~K for H-transfer and at about 200~K for D-transfer. A significant
primary H/D KIE of about 200 is found at 100~K and of 300--600 at 50~K.

At 80--50~K the reaction rate constants of the H-transfer reaction become
almost temperature-independent due to atom tunneling.  Our results clearly
indicate that the title reaction may well be relevant for processes in the
interstellar medium at even lower temperature, even including deuterium.

\begin{acknowledgments}
Alexander Denzel is acknowledged for performing the initial calculations.
This work was financially supported by the German Research Foundation (DFG)
within the Cluster of Excellence in Simulation Technology (EXC 310/2) at the
University of Stuttgart.
\end{acknowledgments}


\begin{thebibliography}{100}%
\makeatletter
\providecommand \@ifxundefined [1]{%
 \@ifx{#1\undefined}
}%
\providecommand \@ifnum [1]{%
 \ifnum #1\expandafter \@firstoftwo
 \else \expandafter \@secondoftwo
 \fi
}%
\providecommand \@ifx [1]{%
 \ifx #1\expandafter \@firstoftwo
 \else \expandafter \@secondoftwo
 \fi
}%
\providecommand \natexlab [1]{#1}%
\providecommand \enquote  [1]{``#1''}%
\providecommand \bibnamefont  [1]{#1}%
\providecommand \bibfnamefont [1]{#1}%
\providecommand \citenamefont [1]{#1}%
\providecommand \href@noop [0]{\@secondoftwo}%
\providecommand \href [0]{\begingroup \@sanitize@url \@href}%
\providecommand \@href[1]{\@@startlink{#1}\@@href}%
\providecommand \@@href[1]{\endgroup#1\@@endlink}%
\providecommand \@sanitize@url [0]{\catcode `\\12\catcode `\$12\catcode
  `\&12\catcode `\#12\catcode `\^12\catcode `\_12\catcode `\%12\relax}%
\providecommand \@@startlink[1]{}%
\providecommand \@@endlink[0]{}%
\providecommand \url  [0]{\begingroup\@sanitize@url \@url }%
\providecommand \@url [1]{\endgroup\@href {#1}{\urlprefix }}%
\providecommand \urlprefix  [0]{URL }%
\providecommand \Eprint [0]{\href }%
\providecommand \doibase [0]{http://dx.doi.org/}%
\providecommand \selectlanguage [0]{\@gobble}%
\providecommand \bibinfo  [0]{\@secondoftwo}%
\providecommand \bibfield  [0]{\@secondoftwo}%
\providecommand \translation [1]{[#1]}%
\providecommand \BibitemOpen [0]{}%
\providecommand \bibitemStop [0]{}%
\providecommand \bibitemNoStop [0]{.\EOS\space}%
\providecommand \EOS [0]{\spacefactor3000\relax}%
\providecommand \BibitemShut  [1]{\csname bibitem#1\endcsname}%
\let\auto@bib@innerbib\@empty
\bibitem [{\citenamefont {Pieterse}, \citenamefont {Krol},\ and\ \citenamefont
  {R\"ockmann}(2009)}]{pie09}%
  \BibitemOpen
  \bibfield  {author} {\bibinfo {author} {\bibfnamefont {G.}~\bibnamefont
  {Pieterse}}, \bibinfo {author} {\bibfnamefont {M.~C.}\ \bibnamefont {Krol}},
  \ and\ \bibinfo {author} {\bibfnamefont {T.}~\bibnamefont {R\"ockmann}},\
  }\href {\doibase 10.5194/acp-9-8503-2009} {\bibfield  {journal} {\bibinfo
  {journal} {Atmos. Chem. Phys.}\ }\textbf {\bibinfo {volume} {9}},\ \bibinfo
  {pages} {8503} (\bibinfo {year} {2009})}\BibitemShut {NoStop}%
\bibitem [{\citenamefont {Rahn}\ \emph {et~al.}(2003)\citenamefont {Rahn},
  \citenamefont {Eiler}, \citenamefont {Boering}, \citenamefont {Wennberg},
  \citenamefont {McCarthy}, \citenamefont {Tyler}, \citenamefont {Schauffler},
  \citenamefont {Donnelly},\ and\ \citenamefont {Atlas}}]{rah03}%
  \BibitemOpen
  \bibfield  {author} {\bibinfo {author} {\bibfnamefont {T.}~\bibnamefont
  {Rahn}}, \bibinfo {author} {\bibfnamefont {J.~M.}\ \bibnamefont {Eiler}},
  \bibinfo {author} {\bibfnamefont {K.~A.}\ \bibnamefont {Boering}}, \bibinfo
  {author} {\bibfnamefont {P.~O.}\ \bibnamefont {Wennberg}}, \bibinfo {author}
  {\bibfnamefont {M.~C.}\ \bibnamefont {McCarthy}}, \bibinfo {author}
  {\bibfnamefont {S.}~\bibnamefont {Tyler}}, \bibinfo {author} {\bibfnamefont
  {S.}~\bibnamefont {Schauffler}}, \bibinfo {author} {\bibfnamefont
  {S.}~\bibnamefont {Donnelly}}, \ and\ \bibinfo {author} {\bibfnamefont
  {E.}~\bibnamefont {Atlas}},\ }\href@noop {} {\bibfield  {journal} {\bibinfo
  {journal} {Nature}\ }\textbf {\bibinfo {volume} {424}},\ \bibinfo {pages}
  {918} (\bibinfo {year} {2003})}\BibitemShut {NoStop}%
\bibitem [{\citenamefont {Gligorovski}\ \emph {et~al.}(2015)\citenamefont
  {Gligorovski}, \citenamefont {Strekowski}, \citenamefont {Barbati},\ and\
  \citenamefont {Vione}}]{gli15}%
  \BibitemOpen
  \bibfield  {author} {\bibinfo {author} {\bibfnamefont {S.}~\bibnamefont
  {Gligorovski}}, \bibinfo {author} {\bibfnamefont {R.}~\bibnamefont
  {Strekowski}}, \bibinfo {author} {\bibfnamefont {S.}~\bibnamefont {Barbati}},
  \ and\ \bibinfo {author} {\bibfnamefont {D.}~\bibnamefont {Vione}},\ }\href
  {\doibase 10.1021/cr500310b} {\bibfield  {journal} {\bibinfo  {journal}
  {Chem. Rev.}\ }\textbf {\bibinfo {volume} {115}},\ \bibinfo {pages} {13051}
  (\bibinfo {year} {2015})}\BibitemShut {NoStop}%
\bibitem [{\citenamefont {Cuppen}\ and\ \citenamefont {Herbst}(2007)}]{cup07}%
  \BibitemOpen
  \bibfield  {author} {\bibinfo {author} {\bibfnamefont {H.~M.}\ \bibnamefont
  {Cuppen}}\ and\ \bibinfo {author} {\bibfnamefont {E.}~\bibnamefont
  {Herbst}},\ }\href@noop {} {\bibfield  {journal} {\bibinfo  {journal}
  {Astrophys. J.}\ }\textbf {\bibinfo {volume} {668}},\ \bibinfo {pages} {294}
  (\bibinfo {year} {2007})}\BibitemShut {NoStop}%
\bibitem [{\citenamefont {Cuppen}\ \emph {et~al.}(2010)\citenamefont {Cuppen},
  \citenamefont {Ioppolo}, \citenamefont {Romanzin},\ and\ \citenamefont
  {Linnartz}}]{cup10}%
  \BibitemOpen
  \bibfield  {author} {\bibinfo {author} {\bibfnamefont {H.~M.}\ \bibnamefont
  {Cuppen}}, \bibinfo {author} {\bibfnamefont {S.}~\bibnamefont {Ioppolo}},
  \bibinfo {author} {\bibfnamefont {C.}~\bibnamefont {Romanzin}}, \ and\
  \bibinfo {author} {\bibfnamefont {H.}~\bibnamefont {Linnartz}},\ }\href
  {\doibase 10.1039/C0CP00251H} {\bibfield  {journal} {\bibinfo  {journal}
  {Phys. Chem. Chem. Phys.}\ }\textbf {\bibinfo {volume} {12}},\ \bibinfo
  {pages} {12077} (\bibinfo {year} {2010})}\BibitemShut {NoStop}%
\bibitem [{\citenamefont {Lamberts}\ \emph {et~al.}(2013)\citenamefont
  {Lamberts}, \citenamefont {Cuppen}, \citenamefont {Ioppolo},\ and\
  \citenamefont {Linnartz}}]{lam13a}%
  \BibitemOpen
  \bibfield  {author} {\bibinfo {author} {\bibfnamefont {T.}~\bibnamefont
  {Lamberts}}, \bibinfo {author} {\bibfnamefont {H.~M.}\ \bibnamefont
  {Cuppen}}, \bibinfo {author} {\bibfnamefont {S.}~\bibnamefont {Ioppolo}}, \
  and\ \bibinfo {author} {\bibfnamefont {H.}~\bibnamefont {Linnartz}},\ }\href
  {\doibase 10.1039/C3CP00106G} {\bibfield  {journal} {\bibinfo  {journal}
  {Phys. Chem. Chem. Phys.}\ }\textbf {\bibinfo {volume} {15}},\ \bibinfo
  {pages} {8287} (\bibinfo {year} {2013})}\BibitemShut {NoStop}%
\bibitem [{\citenamefont {{Lamberts, T.}}\ \emph {et~al.}(2014)\citenamefont
  {{Lamberts, T.}}, \citenamefont {{Cuppen, H. M.}}, \citenamefont {{Fedoseev,
  G.}}, \citenamefont {{Ioppolo, S.}}, \citenamefont {{Chuang, K.-J.}},\ and\
  \citenamefont {{Linnartz, H.}}}]{lam14}%
  \BibitemOpen
  \bibfield  {author} {\bibinfo {author} {\bibnamefont {{Lamberts, T.}}},
  \bibinfo {author} {\bibnamefont {{Cuppen, H. M.}}}, \bibinfo {author}
  {\bibnamefont {{Fedoseev, G.}}}, \bibinfo {author} {\bibnamefont {{Ioppolo,
  S.}}}, \bibinfo {author} {\bibnamefont {{Chuang, K.-J.}}}, \ and\ \bibinfo
  {author} {\bibnamefont {{Linnartz, H.}}},\ }\href@noop {} {\bibfield
  {journal} {\bibinfo  {journal} {Astron. Astrophys.}\ }\textbf {\bibinfo
  {volume} {570}},\ \bibinfo {pages} {A57} (\bibinfo {year}
  {2014})}\BibitemShut {NoStop}%
\bibitem [{\citenamefont {Oba}\ \emph {et~al.}(2012)\citenamefont {Oba},
  \citenamefont {Watanabe}, \citenamefont {Hama}, \citenamefont {Kuwahata},
  \citenamefont {Hidaka},\ and\ \citenamefont {Kouchi}}]{oba12}%
  \BibitemOpen
  \bibfield  {author} {\bibinfo {author} {\bibfnamefont {Y.}~\bibnamefont
  {Oba}}, \bibinfo {author} {\bibfnamefont {N.}~\bibnamefont {Watanabe}},
  \bibinfo {author} {\bibfnamefont {T.}~\bibnamefont {Hama}}, \bibinfo {author}
  {\bibfnamefont {K.}~\bibnamefont {Kuwahata}}, \bibinfo {author}
  {\bibfnamefont {H.}~\bibnamefont {Hidaka}}, \ and\ \bibinfo {author}
  {\bibfnamefont {A.}~\bibnamefont {Kouchi}},\ }\href@noop {} {\bibfield
  {journal} {\bibinfo  {journal} {Astrophys. J.}\ }\textbf {\bibinfo {volume}
  {749}},\ \bibinfo {pages} {67} (\bibinfo {year} {2012})}\BibitemShut
  {NoStop}%
\bibitem [{\citenamefont {Meisner}\ and\ \citenamefont
  {K\"astner}(2016)}]{mei16}%
  \BibitemOpen
  \bibfield  {author} {\bibinfo {author} {\bibfnamefont {J.}~\bibnamefont
  {Meisner}}\ and\ \bibinfo {author} {\bibfnamefont {J.}~\bibnamefont
  {K\"astner}},\ }\href {\doibase 10.1002/anie.201511028} {\bibfield  {journal}
  {\bibinfo  {journal} {Angew. Chem. Int. Ed.}\ ,\ \bibinfo {pages} {ASAP, DOI
  10.1002/anie.201511028}} (\bibinfo {year} {2016})}\BibitemShut {NoStop}%
\bibitem [{\citenamefont {Bhattacharya}, \citenamefont {Panda},\ and\
  \citenamefont {Meyer}(2010)}]{bha10}%
  \BibitemOpen
  \bibfield  {author} {\bibinfo {author} {\bibfnamefont {S.}~\bibnamefont
  {Bhattacharya}}, \bibinfo {author} {\bibfnamefont {A.~N.}\ \bibnamefont
  {Panda}}, \ and\ \bibinfo {author} {\bibfnamefont {H.-D.}\ \bibnamefont
  {Meyer}},\ }\href {\doibase http://dx.doi.org/10.1063/1.3429609} {\bibfield
  {journal} {\bibinfo  {journal} {J. Chem. Phys.}\ }\textbf {\bibinfo {volume}
  {132}},\ \bibinfo {eid} {214304} (\bibinfo {year} {2010})}\BibitemShut
  {NoStop}%
\bibitem [{\citenamefont {Bhattacharya}, \citenamefont {Panda},\ and\
  \citenamefont {Meyer}(2011)}]{bha11}%
  \BibitemOpen
  \bibfield  {author} {\bibinfo {author} {\bibfnamefont {S.}~\bibnamefont
  {Bhattacharya}}, \bibinfo {author} {\bibfnamefont {A.~N.}\ \bibnamefont
  {Panda}}, \ and\ \bibinfo {author} {\bibfnamefont {H.-D.}\ \bibnamefont
  {Meyer}},\ }\href {\doibase http://dx.doi.org/10.1063/1.3660222} {\bibfield
  {journal} {\bibinfo  {journal} {J. Chem. Phys.}\ }\textbf {\bibinfo {volume}
  {135}},\ \bibinfo {eid} {194302} (\bibinfo {year} {2011})}\BibitemShut
  {NoStop}%
\bibitem [{\citenamefont {Espinosa-Garcia}, \citenamefont {Bonnet},\ and\
  \citenamefont {Corchado}(2010)}]{esp10}%
  \BibitemOpen
  \bibfield  {author} {\bibinfo {author} {\bibfnamefont {J.}~\bibnamefont
  {Espinosa-Garcia}}, \bibinfo {author} {\bibfnamefont {L.}~\bibnamefont
  {Bonnet}}, \ and\ \bibinfo {author} {\bibfnamefont {J.~C.}\ \bibnamefont
  {Corchado}},\ }\href {\doibase 10.1039/B922389D} {\bibfield  {journal}
  {\bibinfo  {journal} {Phys. Chem. Chem. Phys.}\ }\textbf {\bibinfo {volume}
  {12}},\ \bibinfo {pages} {3873} (\bibinfo {year} {2010})}\BibitemShut
  {NoStop}%
\bibitem [{\citenamefont {Nguyen}, \citenamefont {Stanton},\ and\ \citenamefont
  {Barker}(2011)}]{ngu11}%
  \BibitemOpen
  \bibfield  {author} {\bibinfo {author} {\bibfnamefont {T.~L.}\ \bibnamefont
  {Nguyen}}, \bibinfo {author} {\bibfnamefont {J.~F.}\ \bibnamefont {Stanton}},
  \ and\ \bibinfo {author} {\bibfnamefont {J.~R.}\ \bibnamefont {Barker}},\
  }\href {\doibase 10.1021/jp2022743} {\bibfield  {journal} {\bibinfo
  {journal} {J. Phys. Chem. A}\ }\textbf {\bibinfo {volume} {115}},\ \bibinfo
  {pages} {5118} (\bibinfo {year} {2011})}\BibitemShut {NoStop}%
\bibitem [{\citenamefont {Fu}, \citenamefont {Kamarchik},\ and\ \citenamefont
  {Bowman}(2010)}]{fu10}%
  \BibitemOpen
  \bibfield  {author} {\bibinfo {author} {\bibfnamefont {B.}~\bibnamefont
  {Fu}}, \bibinfo {author} {\bibfnamefont {E.}~\bibnamefont {Kamarchik}}, \
  and\ \bibinfo {author} {\bibfnamefont {J.~M.}\ \bibnamefont {Bowman}},\
  }\href {\doibase http://dx.doi.org/10.1063/1.3488167} {\bibfield  {journal}
  {\bibinfo  {journal} {J. Chem. Phys.}\ }\textbf {\bibinfo {volume} {133}},\
  \bibinfo {eid} {164306} (\bibinfo {year} {2010})}\BibitemShut {NoStop}%
\bibitem [{\citenamefont {Fu}\ and\ \citenamefont {Zhang}(2015)}]{fu15}%
  \BibitemOpen
  \bibfield  {author} {\bibinfo {author} {\bibfnamefont {B.}~\bibnamefont
  {Fu}}\ and\ \bibinfo {author} {\bibfnamefont {D.~H.}\ \bibnamefont {Zhang}},\
  }\href {\doibase http://dx.doi.org/10.1063/1.4907918} {\bibfield  {journal}
  {\bibinfo  {journal} {J. Chem. Phys.}\ }\textbf {\bibinfo {volume} {142}},\
  \bibinfo {eid} {064314} (\bibinfo {year} {2015})}\BibitemShut {NoStop}%
\bibitem [{\citenamefont {Chan}\ \emph {et~al.}(2004)\citenamefont {Chan},
  \citenamefont {Bollinger}, \citenamefont {Grewell},\ and\ \citenamefont
  {Dooley}}]{cha04}%
  \BibitemOpen
  \bibfield  {author} {\bibinfo {author} {\bibfnamefont {J.~M.}\ \bibnamefont
  {Chan}}, \bibinfo {author} {\bibfnamefont {J.~A.}\ \bibnamefont {Bollinger}},
  \bibinfo {author} {\bibfnamefont {C.~L.}\ \bibnamefont {Grewell}}, \ and\
  \bibinfo {author} {\bibfnamefont {D.~M.}\ \bibnamefont {Dooley}},\ }\href
  {\doibase 10.1021/ja0398868} {\bibfield  {journal} {\bibinfo  {journal} {J.
  Am. Chem. Soc.}\ }\textbf {\bibinfo {volume} {126}},\ \bibinfo {pages} {3030}
  (\bibinfo {year} {2004})}\BibitemShut {NoStop}%
\bibitem [{\citenamefont {Manthe}\ and\ \citenamefont
  {Matzkies}(2000)}]{man00}%
  \BibitemOpen
  \bibfield  {author} {\bibinfo {author} {\bibfnamefont {U.}~\bibnamefont
  {Manthe}}\ and\ \bibinfo {author} {\bibfnamefont {F.}~\bibnamefont
  {Matzkies}},\ }\href {\doibase http://dx.doi.org/10.1063/1.1290284}
  {\bibfield  {journal} {\bibinfo  {journal} {J. Chem. Phys.}\ }\textbf
  {\bibinfo {volume} {113}},\ \bibinfo {pages} {5725} (\bibinfo {year}
  {2000})}\BibitemShut {NoStop}%
\bibitem [{\citenamefont {Ravishankara}\ \emph {et~al.}(1981)\citenamefont
  {Ravishankara}, \citenamefont {Nicovich}, \citenamefont {Thompson},\ and\
  \citenamefont {Tully}}]{rav81}%
  \BibitemOpen
  \bibfield  {author} {\bibinfo {author} {\bibfnamefont {A.~R.}\ \bibnamefont
  {Ravishankara}}, \bibinfo {author} {\bibfnamefont {J.~M.}\ \bibnamefont
  {Nicovich}}, \bibinfo {author} {\bibfnamefont {R.~L.}\ \bibnamefont
  {Thompson}}, \ and\ \bibinfo {author} {\bibfnamefont {F.~P.}\ \bibnamefont
  {Tully}},\ }\href {\doibase 10.1021/j150617a018} {\bibfield  {journal}
  {\bibinfo  {journal} {J. Phys. Chem.}\ }\textbf {\bibinfo {volume} {85}},\
  \bibinfo {pages} {2498} (\bibinfo {year} {1981})}\BibitemShut {NoStop}%
\bibitem [{\citenamefont {Talukdar}\ \emph {et~al.}(1996)\citenamefont
  {Talukdar}, \citenamefont {Gierczak}, \citenamefont {Goldfarb}, \citenamefont
  {Rudich}, \citenamefont {Rao},\ and\ \citenamefont {Ravishankara}}]{tal96}%
  \BibitemOpen
  \bibfield  {author} {\bibinfo {author} {\bibfnamefont {R.~K.}\ \bibnamefont
  {Talukdar}}, \bibinfo {author} {\bibfnamefont {T.}~\bibnamefont {Gierczak}},
  \bibinfo {author} {\bibfnamefont {L.}~\bibnamefont {Goldfarb}}, \bibinfo
  {author} {\bibfnamefont {Y.}~\bibnamefont {Rudich}}, \bibinfo {author}
  {\bibfnamefont {B.~S.~M.}\ \bibnamefont {Rao}}, \ and\ \bibinfo {author}
  {\bibfnamefont {A.~R.}\ \bibnamefont {Ravishankara}},\ }\href {\doibase
  10.1021/jp9518724} {\bibfield  {journal} {\bibinfo  {journal} {J. Phys.
  Chem.}\ }\textbf {\bibinfo {volume} {100}},\ \bibinfo {pages} {3037}
  (\bibinfo {year} {1996})}\BibitemShut {NoStop}%
\bibitem [{\citenamefont {Krasnoperov}\ and\ \citenamefont
  {Michael}(2004)}]{kra04}%
  \BibitemOpen
  \bibfield  {author} {\bibinfo {author} {\bibfnamefont {L.~N.}\ \bibnamefont
  {Krasnoperov}}\ and\ \bibinfo {author} {\bibfnamefont {J.~V.}\ \bibnamefont
  {Michael}},\ }\href {\doibase 10.1021/jp040186e} {\bibfield  {journal}
  {\bibinfo  {journal} {J. Phys. Chem. A}\ }\textbf {\bibinfo {volume} {108}},\
  \bibinfo {pages} {5643} (\bibinfo {year} {2004})}\BibitemShut {NoStop}%
\bibitem [{\citenamefont {Orkin}\ \emph {et~al.}(2006)\citenamefont {Orkin},
  \citenamefont {Kozlov}, \citenamefont {Poskrebyshev},\ and\ \citenamefont
  {Kurylo}}]{ork06}%
  \BibitemOpen
  \bibfield  {author} {\bibinfo {author} {\bibfnamefont {V.~L.}\ \bibnamefont
  {Orkin}}, \bibinfo {author} {\bibfnamefont {S.~N.}\ \bibnamefont {Kozlov}},
  \bibinfo {author} {\bibfnamefont {G.~A.}\ \bibnamefont {Poskrebyshev}}, \
  and\ \bibinfo {author} {\bibfnamefont {M.~J.}\ \bibnamefont {Kurylo}},\
  }\href {\doibase 10.1021/jp057035b} {\bibfield  {journal} {\bibinfo
  {journal} {J. Phys. Chem. A}\ }\textbf {\bibinfo {volume} {110}},\ \bibinfo
  {pages} {6978} (\bibinfo {year} {2006})}\BibitemShut {NoStop}%
\bibitem [{\citenamefont {Lam}, \citenamefont {Davidson},\ and\ \citenamefont
  {Hanson}(2013)}]{lam13}%
  \BibitemOpen
  \bibfield  {author} {\bibinfo {author} {\bibfnamefont {K.-Y.}\ \bibnamefont
  {Lam}}, \bibinfo {author} {\bibfnamefont {D.~F.}\ \bibnamefont {Davidson}}, \
  and\ \bibinfo {author} {\bibfnamefont {R.~K.}\ \bibnamefont {Hanson}},\
  }\href {\doibase 10.1002/kin.20771} {\bibfield  {journal} {\bibinfo
  {journal} {Int. J. Chem. Kinet.}\ }\textbf {\bibinfo {volume} {45}},\
  \bibinfo {pages} {363} (\bibinfo {year} {2013})}\BibitemShut {NoStop}%
\bibitem [{\citenamefont {Castillo}(2002)}]{cas02}%
  \BibitemOpen
  \bibfield  {author} {\bibinfo {author} {\bibfnamefont {J.~F.}\ \bibnamefont
  {Castillo}},\ }\href {\doibase
  10.1002/1439-7641(20020415)3:4<320::AID-CPHC320>3.0.CO;2-B} {\bibfield
  {journal} {\bibinfo  {journal} {ChemPhysChem}\ }\textbf {\bibinfo {volume}
  {3}},\ \bibinfo {pages} {320} (\bibinfo {year} {2002})}\BibitemShut {NoStop}%
\bibitem [{\citenamefont {Smith}\ and\ \citenamefont
  {Fleming~Crim}(2002)}]{smi02a}%
  \BibitemOpen
  \bibfield  {author} {\bibinfo {author} {\bibfnamefont {I.~W.~M.}\
  \bibnamefont {Smith}}\ and\ \bibinfo {author} {\bibfnamefont
  {F.}~\bibnamefont {Fleming~Crim}},\ }\href {\doibase 10.1039/B200985B}
  {\bibfield  {journal} {\bibinfo  {journal} {Phys. Chem. Chem. Phys.}\
  }\textbf {\bibinfo {volume} {4}},\ \bibinfo {pages} {3543} (\bibinfo {year}
  {2002})}\BibitemShut {NoStop}%
\bibitem [{\citenamefont {Langer}(1967)}]{lan67}%
  \BibitemOpen
  \bibfield  {author} {\bibinfo {author} {\bibfnamefont {J.~S.}\ \bibnamefont
  {Langer}},\ }\href {\doibase 10.1016/0003-4916(67)90200-X} {\bibfield
  {journal} {\bibinfo  {journal} {Ann. Phys. (N.Y.)}\ }\textbf {\bibinfo
  {volume} {41}},\ \bibinfo {pages} {108} (\bibinfo {year} {1967})}\BibitemShut
  {NoStop}%
\bibitem [{\citenamefont {Miller}(1975)}]{mil75}%
  \BibitemOpen
  \bibfield  {author} {\bibinfo {author} {\bibfnamefont {W.~H.}\ \bibnamefont
  {Miller}},\ }\href {\doibase 10.1063/1.430676} {\bibfield  {journal}
  {\bibinfo  {journal} {J. Chem. Phys.}\ }\textbf {\bibinfo {volume} {62}},\
  \bibinfo {pages} {1899} (\bibinfo {year} {1975})}\BibitemShut {NoStop}%
\bibitem [{\citenamefont {Coleman}(1977)}]{col77}%
  \BibitemOpen
  \bibfield  {author} {\bibinfo {author} {\bibfnamefont {S.}~\bibnamefont
  {Coleman}},\ }\href {\doibase 10.1103/PhysRevD.15.2929} {\bibfield  {journal}
  {\bibinfo  {journal} {Phys. Rev. D}\ }\textbf {\bibinfo {volume} {15}},\
  \bibinfo {pages} {2929} (\bibinfo {year} {1977})}\BibitemShut {NoStop}%
\bibitem [{\citenamefont {Callan~Jr.}\ and\ \citenamefont
  {Coleman}(1977)}]{cal77}%
  \BibitemOpen
  \bibfield  {author} {\bibinfo {author} {\bibfnamefont {C.~G.}\ \bibnamefont
  {Callan~Jr.}}\ and\ \bibinfo {author} {\bibfnamefont {S.}~\bibnamefont
  {Coleman}},\ }\href {\doibase 10.1103/PhysRevD.16.1762} {\bibfield  {journal}
  {\bibinfo  {journal} {Phys. Rev. D}\ }\textbf {\bibinfo {volume} {16}},\
  \bibinfo {pages} {1762} (\bibinfo {year} {1977})}\BibitemShut {NoStop}%
\bibitem [{\citenamefont {Gildener}\ and\ \citenamefont
  {Patrascioiu}(1977)}]{gil77}%
  \BibitemOpen
  \bibfield  {author} {\bibinfo {author} {\bibfnamefont {E.}~\bibnamefont
  {Gildener}}\ and\ \bibinfo {author} {\bibfnamefont {A.}~\bibnamefont
  {Patrascioiu}},\ }\href {\doibase 10.1103/PhysRevD.16.423} {\bibfield
  {journal} {\bibinfo  {journal} {Phys. Rev. D}\ }\textbf {\bibinfo {volume}
  {16}},\ \bibinfo {pages} {423} (\bibinfo {year} {1977})}\BibitemShut
  {NoStop}%
\bibitem [{\citenamefont {Liu}\ \emph {et~al.}(1993)\citenamefont {Liu},
  \citenamefont {Lynch}, \citenamefont {Truong}, \citenamefont {Lu},
  \citenamefont {Truhlar},\ and\ \citenamefont {Garrett}}]{liu93a}%
  \BibitemOpen
  \bibfield  {author} {\bibinfo {author} {\bibfnamefont {Y.~P.}\ \bibnamefont
  {Liu}}, \bibinfo {author} {\bibfnamefont {G.~C.}\ \bibnamefont {Lynch}},
  \bibinfo {author} {\bibfnamefont {T.~N.}\ \bibnamefont {Truong}}, \bibinfo
  {author} {\bibfnamefont {D.}~\bibnamefont {Lu}}, \bibinfo {author}
  {\bibfnamefont {D.~G.}\ \bibnamefont {Truhlar}}, \ and\ \bibinfo {author}
  {\bibfnamefont {B.~C.}\ \bibnamefont {Garrett}},\ }\href {\doibase
  10.1021/ja00059a041} {\bibfield  {journal} {\bibinfo  {journal} {J. Am. Chem.
  Soc.}\ }\textbf {\bibinfo {volume} {115}},\ \bibinfo {pages} {2408} (\bibinfo
  {year} {1993})}\BibitemShut {NoStop}%
\bibitem [{\citenamefont {Truhlar}(1994)}]{Truhlar_Faraday1994}%
  \BibitemOpen
  \bibfield  {author} {\bibinfo {author} {\bibfnamefont {D.~G.}\ \bibnamefont
  {Truhlar}},\ }\href {\doibase 10.1039/FT9949001733} {\bibfield  {journal}
  {\bibinfo  {journal} {J. Chem. Soc.{,} Faraday Trans.}\ }\textbf {\bibinfo
  {volume} {90}},\ \bibinfo {pages} {1740} (\bibinfo {year}
  {1994})}\BibitemShut {NoStop}%
\bibitem [{\citenamefont {Affleck}(1981)}]{aff81}%
  \BibitemOpen
  \bibfield  {author} {\bibinfo {author} {\bibfnamefont {I.}~\bibnamefont
  {Affleck}},\ }\href {\doibase 10.1103/PhysRevLett.46.388} {\bibfield
  {journal} {\bibinfo  {journal} {Phys. Rev. Lett.}\ }\textbf {\bibinfo
  {volume} {46}},\ \bibinfo {pages} {388} (\bibinfo {year} {1981})}\BibitemShut
  {NoStop}%
\bibitem [{\citenamefont {Coleman}(1988)}]{col88}%
  \BibitemOpen
  \bibfield  {author} {\bibinfo {author} {\bibfnamefont {S.}~\bibnamefont
  {Coleman}},\ }\href {\doibase 10.1016/0550-3213(88)90308-2} {\bibfield
  {journal} {\bibinfo  {journal} {Nucl. Phys. B}\ }\textbf {\bibinfo {volume}
  {298}},\ \bibinfo {pages} {178} (\bibinfo {year} {1988})}\BibitemShut
  {NoStop}%
\bibitem [{\citenamefont {H\"anggi}, \citenamefont {Talkner},\ and\
  \citenamefont {Borkovec}(1990)}]{han90}%
  \BibitemOpen
  \bibfield  {author} {\bibinfo {author} {\bibfnamefont {P.}~\bibnamefont
  {H\"anggi}}, \bibinfo {author} {\bibfnamefont {P.}~\bibnamefont {Talkner}}, \
  and\ \bibinfo {author} {\bibfnamefont {M.}~\bibnamefont {Borkovec}},\ }\href
  {\doibase 10.1103/RevModPhys.62.251} {\bibfield  {journal} {\bibinfo
  {journal} {Rev. Mod. Phys.}\ }\textbf {\bibinfo {volume} {62}},\ \bibinfo
  {pages} {251} (\bibinfo {year} {1990})}\BibitemShut {NoStop}%
\bibitem [{\citenamefont {Benderskii}, \citenamefont {Makarov},\ and\
  \citenamefont {Wight}(1994)}]{ben94}%
  \BibitemOpen
  \bibfield  {author} {\bibinfo {author} {\bibfnamefont {V.~A.}\ \bibnamefont
  {Benderskii}}, \bibinfo {author} {\bibfnamefont {D.~E.}\ \bibnamefont
  {Makarov}}, \ and\ \bibinfo {author} {\bibfnamefont {C.~A.}\ \bibnamefont
  {Wight}},\ }\href {\doibase 10.1002/9780470141472.ch3} {\bibfield  {journal}
  {\bibinfo  {journal} {Adv. Chem. Phys.}\ }\textbf {\bibinfo {volume} {88}},\
  \bibinfo {pages} {55} (\bibinfo {year} {1994})}\BibitemShut {NoStop}%
\bibitem [{\citenamefont {Messina}, \citenamefont {Schenter},\ and\
  \citenamefont {Garrett}(1995)}]{mes95}%
  \BibitemOpen
  \bibfield  {author} {\bibinfo {author} {\bibfnamefont {M.}~\bibnamefont
  {Messina}}, \bibinfo {author} {\bibfnamefont {G.~K.}\ \bibnamefont
  {Schenter}}, \ and\ \bibinfo {author} {\bibfnamefont {B.~C.}\ \bibnamefont
  {Garrett}},\ }\href {\doibase 10.1063/1.470227} {\bibfield  {journal}
  {\bibinfo  {journal} {J. Chem. Phys.}\ }\textbf {\bibinfo {volume} {103}},\
  \bibinfo {pages} {3430} (\bibinfo {year} {1995})}\BibitemShut {NoStop}%
\bibitem [{\citenamefont {Richardson}\ and\ \citenamefont
  {Althorpe}(2009)}]{ric09}%
  \BibitemOpen
  \bibfield  {author} {\bibinfo {author} {\bibfnamefont {J.~O.}\ \bibnamefont
  {Richardson}}\ and\ \bibinfo {author} {\bibfnamefont {S.~C.}\ \bibnamefont
  {Althorpe}},\ }\href {\doibase 10.1063/1.3267318} {\bibfield  {journal}
  {\bibinfo  {journal} {J. Chem. Phys.}\ }\textbf {\bibinfo {volume} {131}},\
  \bibinfo {pages} {214106} (\bibinfo {year} {2009})}\BibitemShut {NoStop}%
\bibitem [{\citenamefont {Althorpe}(2011)}]{alt11}%
  \BibitemOpen
  \bibfield  {author} {\bibinfo {author} {\bibfnamefont {S.~C.}\ \bibnamefont
  {Althorpe}},\ }\href {\doibase 10.1063/1.3563045} {\bibfield  {journal}
  {\bibinfo  {journal} {J. Chem. Phys.}\ }\textbf {\bibinfo {volume} {134}},\
  \bibinfo {pages} {114104} (\bibinfo {year} {2011})}\BibitemShut {NoStop}%
\bibitem [{\citenamefont {Rommel}, \citenamefont {Goumans},\ and\ \citenamefont
  {K\"astner}(2011)}]{rom11}%
  \BibitemOpen
  \bibfield  {author} {\bibinfo {author} {\bibfnamefont {J.~B.}\ \bibnamefont
  {Rommel}}, \bibinfo {author} {\bibfnamefont {T.~P.~M.}\ \bibnamefont
  {Goumans}}, \ and\ \bibinfo {author} {\bibfnamefont {J.}~\bibnamefont
  {K\"astner}},\ }\href {\doibase 10.1021/ct100658y} {\bibfield  {journal}
  {\bibinfo  {journal} {J. Chem. Theory Comput.}\ }\textbf {\bibinfo {volume}
  {7}},\ \bibinfo {pages} {690} (\bibinfo {year} {2011})}\BibitemShut {NoStop}%
\bibitem [{\citenamefont {Rommel}\ and\ \citenamefont
  {K\"astner}(2011)}]{rom11b}%
  \BibitemOpen
  \bibfield  {author} {\bibinfo {author} {\bibfnamefont {J.~B.}\ \bibnamefont
  {Rommel}}\ and\ \bibinfo {author} {\bibfnamefont {J.}~\bibnamefont
  {K\"astner}},\ }\href {\doibase 10.1063/1.3587240} {\bibfield  {journal}
  {\bibinfo  {journal} {J. Chem. Phys.}\ }\textbf {\bibinfo {volume} {134}},\
  \bibinfo {pages} {184107} (\bibinfo {year} {2011})}\BibitemShut {NoStop}%
\bibitem [{\citenamefont {Richardson}(2016)}]{ric16}%
  \BibitemOpen
  \bibfield  {author} {\bibinfo {author} {\bibfnamefont {J.~O.}\ \bibnamefont
  {Richardson}},\ }\href@noop {} {\bibfield  {journal} {\bibinfo  {journal}
  {J. Chem. Phys.}\ }\textbf {\bibinfo {volume} {144}},\
  \bibinfo {eid} {114106} (\bibinfo {year} {2016})}\BibitemShut {NoStop}%
\bibitem [{\citenamefont {Feynman}(1948)}]{fey48}%
  \BibitemOpen
  \bibfield  {author} {\bibinfo {author} {\bibfnamefont {R.~P.}\ \bibnamefont
  {Feynman}},\ }\href {\doibase 10.1103/RevModPhys.20.367} {\bibfield
  {journal} {\bibinfo  {journal} {Rev. Mod. Phys.}\ }\textbf {\bibinfo {volume}
  {20}},\ \bibinfo {pages} {367} (\bibinfo {year} {1948})}\BibitemShut
  {NoStop}%
\bibitem [{\citenamefont {K\"astner}(2014)}]{kae14}%
  \BibitemOpen
  \bibfield  {author} {\bibinfo {author} {\bibfnamefont {J.}~\bibnamefont
  {K\"astner}},\ }\href {\doibase 10.1002/wcms.1165} {\bibfield  {journal}
  {\bibinfo  {journal} {WIREs Comput. Mol. Sci.}\ }\textbf {\bibinfo {volume}
  {4}},\ \bibinfo {pages} {158} (\bibinfo {year} {2014})}\BibitemShut {NoStop}%
\bibitem [{\citenamefont {Chapman}, \citenamefont {Garrett},\ and\
  \citenamefont {Miller}(1975)}]{cha75}%
  \BibitemOpen
  \bibfield  {author} {\bibinfo {author} {\bibfnamefont {S.}~\bibnamefont
  {Chapman}}, \bibinfo {author} {\bibfnamefont {B.~C.}\ \bibnamefont
  {Garrett}}, \ and\ \bibinfo {author} {\bibfnamefont {W.~H.}\ \bibnamefont
  {Miller}},\ }\href {\doibase 10.1063/1.431620} {\bibfield  {journal}
  {\bibinfo  {journal} {J. Chem. Phys.}\ }\textbf {\bibinfo {volume} {63}},\
  \bibinfo {pages} {2710} (\bibinfo {year} {1975})}\BibitemShut {NoStop}%
\bibitem [{\citenamefont {Mills}\ and\ \citenamefont
  {J{\'o}nsson}(1994)}]{mil94}%
  \BibitemOpen
  \bibfield  {author} {\bibinfo {author} {\bibfnamefont {G.}~\bibnamefont
  {Mills}}\ and\ \bibinfo {author} {\bibfnamefont {H.}~\bibnamefont
  {J{\'o}nsson}},\ }\href {\doibase 10.1103/PhysRevLett.72.1124} {\bibfield
  {journal} {\bibinfo  {journal} {Phys. Rev. Lett.}\ }\textbf {\bibinfo
  {volume} {72}},\ \bibinfo {pages} {1124} (\bibinfo {year}
  {1994})}\BibitemShut {NoStop}%
\bibitem [{\citenamefont {Mills}, \citenamefont {J{\'o}nsson},\ and\
  \citenamefont {Schenter}(1995)}]{mil95}%
  \BibitemOpen
  \bibfield  {author} {\bibinfo {author} {\bibfnamefont {G.}~\bibnamefont
  {Mills}}, \bibinfo {author} {\bibfnamefont {H.}~\bibnamefont {J{\'o}nsson}},
  \ and\ \bibinfo {author} {\bibfnamefont {G.~K.}\ \bibnamefont {Schenter}},\
  }\href {\doibase 10.1016/0039-6028(94)00731-4} {\bibfield  {journal}
  {\bibinfo  {journal} {Surf. Sci.}\ }\textbf {\bibinfo {volume} {324}},\
  \bibinfo {pages} {305} (\bibinfo {year} {1995})}\BibitemShut {NoStop}%
\bibitem [{\citenamefont {Mills}\ \emph {et~al.}(1997)\citenamefont {Mills},
  \citenamefont {Schenter}, \citenamefont {Makarov},\ and\ \citenamefont
  {J{\'o}nsson}}]{mil97}%
  \BibitemOpen
  \bibfield  {author} {\bibinfo {author} {\bibfnamefont {G.}~\bibnamefont
  {Mills}}, \bibinfo {author} {\bibfnamefont {G.~K.}\ \bibnamefont {Schenter}},
  \bibinfo {author} {\bibfnamefont {D.~E.}\ \bibnamefont {Makarov}}, \ and\
  \bibinfo {author} {\bibfnamefont {H.}~\bibnamefont {J{\'o}nsson}},\ }\href
  {\doibase 10.1016/S0009-2614(97)00886-5} {\bibfield  {journal} {\bibinfo
  {journal} {Chem. Phys. Lett.}\ }\textbf {\bibinfo {volume} {278}},\ \bibinfo
  {pages} {91} (\bibinfo {year} {1997})}\BibitemShut {NoStop}%
\bibitem [{\citenamefont {Siebrand}\ \emph {et~al.}(1999)\citenamefont
  {Siebrand}, \citenamefont {Smedarchina}, \citenamefont {Zgierski},\ and\
  \citenamefont {Fern\'andez-Ramos}}]{sie99}%
  \BibitemOpen
  \bibfield  {author} {\bibinfo {author} {\bibfnamefont {W.}~\bibnamefont
  {Siebrand}}, \bibinfo {author} {\bibfnamefont {Z.}~\bibnamefont
  {Smedarchina}}, \bibinfo {author} {\bibfnamefont {M.~Z.}\ \bibnamefont
  {Zgierski}}, \ and\ \bibinfo {author} {\bibfnamefont {A.}~\bibnamefont
  {Fern\'andez-Ramos}},\ }\href {\doibase 10.1080/014423599229992} {\bibfield
  {journal} {\bibinfo  {journal} {Int. Rev. Phys. Chem.}\ }\textbf {\bibinfo
  {volume} {18}},\ \bibinfo {pages} {5} (\bibinfo {year} {1999})}\BibitemShut
  {NoStop}%
\bibitem [{\citenamefont {Smedarchina}\ \emph {et~al.}(2003)\citenamefont
  {Smedarchina}, \citenamefont {Siebrand}, \citenamefont {Fern\'andez-Ramos},\
  and\ \citenamefont {Cui}}]{sme03}%
  \BibitemOpen
  \bibfield  {author} {\bibinfo {author} {\bibfnamefont {Z.}~\bibnamefont
  {Smedarchina}}, \bibinfo {author} {\bibfnamefont {W.}~\bibnamefont
  {Siebrand}}, \bibinfo {author} {\bibfnamefont {A.}~\bibnamefont
  {Fern\'andez-Ramos}}, \ and\ \bibinfo {author} {\bibfnamefont
  {Q.}~\bibnamefont {Cui}},\ }\href {\doibase 10.1021/ja0210594} {\bibfield
  {journal} {\bibinfo  {journal} {J. Am. Chem. Soc.}\ }\textbf {\bibinfo
  {volume} {125}},\ \bibinfo {pages} {243} (\bibinfo {year}
  {2003})}\BibitemShut {NoStop}%
\bibitem [{\citenamefont {Qian}\ \emph {et~al.}(2007)\citenamefont {Qian},
  \citenamefont {Ren}, \citenamefont {Shi}, \citenamefont {E},\ and\
  \citenamefont {Shen}}]{qia07}%
  \BibitemOpen
  \bibfield  {author} {\bibinfo {author} {\bibfnamefont {T.}~\bibnamefont
  {Qian}}, \bibinfo {author} {\bibfnamefont {W.}~\bibnamefont {Ren}}, \bibinfo
  {author} {\bibfnamefont {J.}~\bibnamefont {Shi}}, \bibinfo {author}
  {\bibfnamefont {W.}~\bibnamefont {E}}, \ and\ \bibinfo {author}
  {\bibfnamefont {P.}~\bibnamefont {Shen}},\ }\href {\doibase
  10.1016/j.physa.2007.01.005} {\bibfield  {journal} {\bibinfo  {journal}
  {Physica A}\ }\textbf {\bibinfo {volume} {379}},\ \bibinfo {pages} {491}
  (\bibinfo {year} {2007})}\BibitemShut {NoStop}%
\bibitem [{\citenamefont {Andersson}\ \emph {et~al.}(2009)\citenamefont
  {Andersson}, \citenamefont {Nyman}, \citenamefont {Arnaldsson}, \citenamefont
  {Manthe},\ and\ \citenamefont {J{\'o}nsson}}]{and09}%
  \BibitemOpen
  \bibfield  {author} {\bibinfo {author} {\bibfnamefont {S.}~\bibnamefont
  {Andersson}}, \bibinfo {author} {\bibfnamefont {G.}~\bibnamefont {Nyman}},
  \bibinfo {author} {\bibfnamefont {A.}~\bibnamefont {Arnaldsson}}, \bibinfo
  {author} {\bibfnamefont {U.}~\bibnamefont {Manthe}}, \ and\ \bibinfo {author}
  {\bibfnamefont {H.}~\bibnamefont {J{\'o}nsson}},\ }\href {\doibase
  10.1021/jp811070w} {\bibfield  {journal} {\bibinfo  {journal} {J. Phys. Chem.
  A}\ }\textbf {\bibinfo {volume} {113}},\ \bibinfo {pages} {4468} (\bibinfo
  {year} {2009})}\BibitemShut {NoStop}%
\bibitem [{\citenamefont {Goumans}\ and\ \citenamefont
  {Andersson}(2010)}]{gou10a}%
  \BibitemOpen
  \bibfield  {author} {\bibinfo {author} {\bibfnamefont {T.~P.~M.}\
  \bibnamefont {Goumans}}\ and\ \bibinfo {author} {\bibfnamefont
  {S.}~\bibnamefont {Andersson}},\ }\href {\doibase
  10.1111/j.1365-2966.2010.16836.x} {\bibfield  {journal} {\bibinfo  {journal}
  {Mon. Not. R. Astron. Soc.}\ }\textbf {\bibinfo {volume} {406}},\ \bibinfo
  {pages} {2213} (\bibinfo {year} {2010})}\BibitemShut {NoStop}%
\bibitem [{\citenamefont {Goumans}(2011{\natexlab{a}})}]{gou11}%
  \BibitemOpen
  \bibfield  {author} {\bibinfo {author} {\bibfnamefont {T.~P.~M.}\
  \bibnamefont {Goumans}},\ }\href {\doibase 10.1111/j.1365-2966.2011.18924.x}
  {\bibfield  {journal} {\bibinfo  {journal} {Mon. Not. Roy. Astron. Soc.}\
  }\textbf {\bibinfo {volume} {415}},\ \bibinfo {pages} {3129} (\bibinfo {year}
  {2011}{\natexlab{a}})}\BibitemShut {NoStop}%
\bibitem [{\citenamefont {Goumans}(2011{\natexlab{b}})}]{gou11b}%
  \BibitemOpen
  \bibfield  {author} {\bibinfo {author} {\bibfnamefont {T.~P.~M.}\
  \bibnamefont {Goumans}},\ }\href {\doibase 10.1111/j.1365-2966.2011.18329.x}
  {\bibfield  {journal} {\bibinfo  {journal} {Mon. Not. Roy. Astron. Soc.}\
  }\textbf {\bibinfo {volume} {413}},\ \bibinfo {pages} {26150} (\bibinfo
  {year} {2011}{\natexlab{b}})}\BibitemShut {NoStop}%
\bibitem [{\citenamefont {Goumans}\ and\ \citenamefont
  {K\"astner}(2010)}]{gou10}%
  \BibitemOpen
  \bibfield  {author} {\bibinfo {author} {\bibfnamefont {T.~P.~M.}\
  \bibnamefont {Goumans}}\ and\ \bibinfo {author} {\bibfnamefont
  {J.}~\bibnamefont {K\"astner}},\ }\href {\doibase 10.1002/anie.201001311}
  {\bibfield  {journal} {\bibinfo  {journal} {Angew. Chem., Int. Ed.}\ }\textbf
  {\bibinfo {volume} {49}},\ \bibinfo {pages} {7350} (\bibinfo {year}
  {2010})}\BibitemShut {NoStop}%
\bibitem [{\citenamefont {J\'o{}nsson}(2010)}]{jon10}%
  \BibitemOpen
  \bibfield  {author} {\bibinfo {author} {\bibfnamefont {H.}~\bibnamefont
  {J\'o{}nsson}},\ }\href {\doibase 10.1073/pnas.1006670108} {\bibfield
  {journal} {\bibinfo  {journal} {Proc. Nat. Acad. Sci. U.S.A.}\ }\textbf
  {\bibinfo {volume} {108}},\ \bibinfo {pages} {944} (\bibinfo {year}
  {2010})}\BibitemShut {NoStop}%
\bibitem [{\citenamefont {Meisner}, \citenamefont {Rommel},\ and\ \citenamefont
  {K\"astner}(2011)}]{mei11}%
  \BibitemOpen
  \bibfield  {author} {\bibinfo {author} {\bibfnamefont {J.}~\bibnamefont
  {Meisner}}, \bibinfo {author} {\bibfnamefont {J.~B.}\ \bibnamefont {Rommel}},
  \ and\ \bibinfo {author} {\bibfnamefont {J.}~\bibnamefont {K\"astner}},\
  }\href {\doibase 10.1002/jcc.21930} {\bibfield  {journal} {\bibinfo
  {journal} {J. Comput. Chem.}\ }\textbf {\bibinfo {volume} {32}},\ \bibinfo
  {pages} {3456} (\bibinfo {year} {2011})}\BibitemShut {NoStop}%
\bibitem [{\citenamefont {Goumans}\ and\ \citenamefont
  {K\"astner}(2011)}]{gou11a}%
  \BibitemOpen
  \bibfield  {author} {\bibinfo {author} {\bibfnamefont {T.~P.~M.}\
  \bibnamefont {Goumans}}\ and\ \bibinfo {author} {\bibfnamefont
  {J.}~\bibnamefont {K\"astner}},\ }\href {\doibase 10.1021/jp206048f}
  {\bibfield  {journal} {\bibinfo  {journal} {J. Phys. Chem. A}\ }\textbf
  {\bibinfo {volume} {115}},\ \bibinfo {pages} {10767} (\bibinfo {year}
  {2011})}\BibitemShut {NoStop}%
\bibitem [{\citenamefont {Einarsd\'o{}ttir}\ \emph {et~al.}(2012)\citenamefont
  {Einarsd\'o{}ttir}, \citenamefont {Arnaldsson}, \citenamefont
  {\'O{}skarsson},\ and\ \citenamefont {J\'o{}nsson}}]{ein11}%
  \BibitemOpen
  \bibfield  {author} {\bibinfo {author} {\bibfnamefont {D.~M.}\ \bibnamefont
  {Einarsd\'o{}ttir}}, \bibinfo {author} {\bibfnamefont {A.}~\bibnamefont
  {Arnaldsson}}, \bibinfo {author} {\bibfnamefont {F.}~\bibnamefont
  {\'O{}skarsson}}, \ and\ \bibinfo {author} {\bibfnamefont {H.}~\bibnamefont
  {J\'o{}nsson}},\ }\href {\doibase 10.1007/978-3-642-28145-7_5} {\bibfield
  {journal} {\bibinfo  {journal} {Lect. Notes Comput. Sci.}\ }\textbf {\bibinfo
  {volume} {7134}},\ \bibinfo {pages} {45} (\bibinfo {year}
  {2012})}\BibitemShut {NoStop}%
\bibitem [{\citenamefont {Rommel}\ \emph {et~al.}(2012)\citenamefont {Rommel},
  \citenamefont {Liu}, \citenamefont {Werner},\ and\ \citenamefont
  {K\"astner}}]{rom12}%
  \BibitemOpen
  \bibfield  {author} {\bibinfo {author} {\bibfnamefont {J.~B.}\ \bibnamefont
  {Rommel}}, \bibinfo {author} {\bibfnamefont {Y.}~\bibnamefont {Liu}},
  \bibinfo {author} {\bibfnamefont {H.-J.}\ \bibnamefont {Werner}}, \ and\
  \bibinfo {author} {\bibfnamefont {J.}~\bibnamefont {K\"astner}},\ }\href
  {\doibase 10.1021/jp308526t} {\bibfield  {journal} {\bibinfo  {journal} {J.
  Phys. Chem. B}\ }\textbf {\bibinfo {volume} {116}},\ \bibinfo {pages} {13682}
  (\bibinfo {year} {2012})}\BibitemShut {NoStop}%
\bibitem [{\citenamefont {Kryvohuz}\ and\ \citenamefont
  {Marcus}(2012)}]{kry12}%
  \BibitemOpen
  \bibfield  {author} {\bibinfo {author} {\bibfnamefont {M.}~\bibnamefont
  {Kryvohuz}}\ and\ \bibinfo {author} {\bibfnamefont {R.~A.}\ \bibnamefont
  {Marcus}},\ }\href {\doibase http://dx.doi.org/10.1063/1.4754660} {\bibfield
  {journal} {\bibinfo  {journal} {J. Chem. Phys.}\ }\textbf {\bibinfo {volume}
  {137}},\ \bibinfo {eid} {134107} (\bibinfo {year} {2012})}\BibitemShut
  {NoStop}%
\bibitem [{\citenamefont {K\"astner}(2013)}]{kae13}%
  \BibitemOpen
  \bibfield  {author} {\bibinfo {author} {\bibfnamefont {J.}~\bibnamefont
  {K\"astner}},\ }\href {\doibase 10.1002/chem.201203651} {\bibfield  {journal}
  {\bibinfo  {journal} {Chem. Eur. J.}\ }\textbf {\bibinfo {volume} {19}},\
  \bibinfo {pages} {8207} (\bibinfo {year} {2013})}\BibitemShut {NoStop}%
\bibitem [{\citenamefont {\'Alvarez-Barcia}, \citenamefont {Flores},\ and\
  \citenamefont {K\"astner}(2014)}]{alv14}%
  \BibitemOpen
  \bibfield  {author} {\bibinfo {author} {\bibfnamefont {S.}~\bibnamefont
  {\'Alvarez-Barcia}}, \bibinfo {author} {\bibfnamefont {J.~R.}\ \bibnamefont
  {Flores}}, \ and\ \bibinfo {author} {\bibfnamefont {J.}~\bibnamefont
  {K\"astner}},\ }\href {\doibase 10.1021/jp411189m} {\bibfield  {journal}
  {\bibinfo  {journal} {J. Phys. Chem. A}\ }\textbf {\bibinfo {volume} {118}},\
  \bibinfo {pages} {78} (\bibinfo {year} {2014})}\BibitemShut {NoStop}%
\bibitem [{\citenamefont {Kryvohuz}(2014)}]{kry14}%
  \BibitemOpen
  \bibfield  {author} {\bibinfo {author} {\bibfnamefont {M.}~\bibnamefont
  {Kryvohuz}},\ }\href {\doibase 10.1021/jp4099073} {\bibfield  {journal}
  {\bibinfo  {journal} {J. Phys. Chem. A}\ }\textbf {\bibinfo {volume} {118}},\
  \bibinfo {pages} {535} (\bibinfo {year} {2014})}\BibitemShut {NoStop}%
\bibitem [{\citenamefont {Garrett}\ \emph {et~al.}(1980)\citenamefont
  {Garrett}, \citenamefont {Truhlar}, \citenamefont {Grev},\ and\ \citenamefont
  {Magnuson}}]{gar80}%
  \BibitemOpen
  \bibfield  {author} {\bibinfo {author} {\bibfnamefont {B.~C.}\ \bibnamefont
  {Garrett}}, \bibinfo {author} {\bibfnamefont {D.~G.}\ \bibnamefont
  {Truhlar}}, \bibinfo {author} {\bibfnamefont {R.~S.}\ \bibnamefont {Grev}}, \
  and\ \bibinfo {author} {\bibfnamefont {A.~W.}\ \bibnamefont {Magnuson}},\
  }\href {\doibase 10.1021/j100450a013} {\bibfield  {journal} {\bibinfo
  {journal} {J. Phys. Chem.}\ }\textbf {\bibinfo {volume} {84}},\ \bibinfo
  {pages} {1730} (\bibinfo {year} {1980})}\BibitemShut {NoStop}%
\bibitem [{\citenamefont {Truhlar}\ \emph {et~al.}(1983)\citenamefont
  {Truhlar}, \citenamefont {Issacson}, \citenamefont {Skodje},\ and\
  \citenamefont {Garrett}}]{tru83}%
  \BibitemOpen
  \bibfield  {author} {\bibinfo {author} {\bibfnamefont {D.}~\bibnamefont
  {Truhlar}}, \bibinfo {author} {\bibfnamefont {A.}~\bibnamefont {Issacson}},
  \bibinfo {author} {\bibfnamefont {R.}~\bibnamefont {Skodje}}, \ and\ \bibinfo
  {author} {\bibfnamefont {B.}~\bibnamefont {Garrett}},\ }\href {\doibase
  10.1021/j100245a604} {\bibfield  {journal} {\bibinfo  {journal} {J. Phys.
  Chem.}\ }\textbf {\bibinfo {volume} {87}},\ \bibinfo {pages} {4554} (\bibinfo
  {year} {1983})}\BibitemShut {NoStop}%
\bibitem [{\citenamefont {Skodje}, \citenamefont {Truhlar},\ and\ \citenamefont
  {Garrett}(1981)}]{sko81}%
  \BibitemOpen
  \bibfield  {author} {\bibinfo {author} {\bibfnamefont {R.~T.}\ \bibnamefont
  {Skodje}}, \bibinfo {author} {\bibfnamefont {D.~G.}\ \bibnamefont {Truhlar}},
  \ and\ \bibinfo {author} {\bibfnamefont {B.~C.}\ \bibnamefont {Garrett}},\
  }\href {\doibase 10.1021/j150621a001} {\bibfield  {journal} {\bibinfo
  {journal} {J. Phys. Chem.}\ }\textbf {\bibinfo {volume} {85}},\ \bibinfo
  {pages} {3019} (\bibinfo {year} {1981})}\BibitemShut {NoStop}%
\bibitem [{\citenamefont {Garrett}\ \emph {et~al.}(1983)\citenamefont
  {Garrett}, \citenamefont {Truhlar}, \citenamefont {Wagner},\ and\
  \citenamefont {Dunning~Jr.}}]{gar83}%
  \BibitemOpen
  \bibfield  {author} {\bibinfo {author} {\bibfnamefont {B.~C.}\ \bibnamefont
  {Garrett}}, \bibinfo {author} {\bibfnamefont {D.~G.}\ \bibnamefont
  {Truhlar}}, \bibinfo {author} {\bibfnamefont {A.~F.}\ \bibnamefont {Wagner}},
  \ and\ \bibinfo {author} {\bibfnamefont {T.~H.}\ \bibnamefont
  {Dunning~Jr.}},\ }\href {\doibase 10.1063/1.445323} {\bibfield  {journal}
  {\bibinfo  {journal} {J. Chem. Phys.}\ }\textbf {\bibinfo {volume} {78}},\
  \bibinfo {pages} {4400} (\bibinfo {year} {1983})}\BibitemShut {NoStop}%
\bibitem [{\citenamefont {Garrett}\ \emph {et~al.}(1985)\citenamefont
  {Garrett}, \citenamefont {Abusalbi}, \citenamefont {Kouri},\ and\
  \citenamefont {Truhlar}}]{gar85}%
  \BibitemOpen
  \bibfield  {author} {\bibinfo {author} {\bibfnamefont {B.~C.}\ \bibnamefont
  {Garrett}}, \bibinfo {author} {\bibfnamefont {N.}~\bibnamefont {Abusalbi}},
  \bibinfo {author} {\bibfnamefont {D.~J.}\ \bibnamefont {Kouri}}, \ and\
  \bibinfo {author} {\bibfnamefont {D.~G.}\ \bibnamefont {Truhlar}},\ }\href
  {\doibase 10.1063/1.449318} {\bibfield  {journal} {\bibinfo  {journal} {J.
  Chem. Phys.}\ }\textbf {\bibinfo {volume} {83}},\ \bibinfo {pages} {2252}
  (\bibinfo {year} {1985})}\BibitemShut {NoStop}%
\bibitem [{\citenamefont {Fernandez-Ramos}\ and\ \citenamefont
  {Truhlar}(2001)}]{fer01}%
  \BibitemOpen
  \bibfield  {author} {\bibinfo {author} {\bibfnamefont {A.}~\bibnamefont
  {Fernandez-Ramos}}\ and\ \bibinfo {author} {\bibfnamefont {D.~G.}\
  \bibnamefont {Truhlar}},\ }\href {\doibase
  http://dx.doi.org/10.1063/1.1329893} {\bibfield  {journal} {\bibinfo
  {journal} {J. Chem. Phys.}\ }\textbf {\bibinfo {volume} {114}},\ \bibinfo
  {pages} {1491} (\bibinfo {year} {2001})}\BibitemShut {NoStop}%
\bibitem [{\citenamefont {Fern\'andez-Ramos}\ \emph {et~al.}(2006)\citenamefont
  {Fern\'andez-Ramos}, \citenamefont {Miller}, \citenamefont {Klippenstein},\
  and\ \citenamefont {Truhlar}}]{fer06}%
  \BibitemOpen
  \bibfield  {author} {\bibinfo {author} {\bibfnamefont {A.}~\bibnamefont
  {Fern\'andez-Ramos}}, \bibinfo {author} {\bibfnamefont {J.~A.}\ \bibnamefont
  {Miller}}, \bibinfo {author} {\bibfnamefont {S.~J.}\ \bibnamefont
  {Klippenstein}}, \ and\ \bibinfo {author} {\bibfnamefont {D.~G.}\
  \bibnamefont {Truhlar}},\ }\href {\doibase 10.1021/cr050205w} {\bibfield
  {journal} {\bibinfo  {journal} {Chem. Rev.}\ }\textbf {\bibinfo {volume}
  {106}},\ \bibinfo {pages} {4518} (\bibinfo {year} {2006})}\BibitemShut
  {NoStop}%
\bibitem [{\citenamefont {Fern\'andez-Ramos}\ \emph {et~al.}(2007)\citenamefont
  {Fern\'andez-Ramos}, \citenamefont {Ellingson}, \citenamefont {Garrett},\
  and\ \citenamefont {Truhlar}}]{fer07}%
  \BibitemOpen
  \bibfield  {author} {\bibinfo {author} {\bibfnamefont {A.}~\bibnamefont
  {Fern\'andez-Ramos}}, \bibinfo {author} {\bibfnamefont {B.~A.}\ \bibnamefont
  {Ellingson}}, \bibinfo {author} {\bibfnamefont {B.~C.}\ \bibnamefont
  {Garrett}}, \ and\ \bibinfo {author} {\bibfnamefont {D.~G.}\ \bibnamefont
  {Truhlar}},\ }\enquote {\bibinfo {title} {Reviews in computational
  chemistry},}\ \ (\bibinfo  {publisher} {Wiley-VCH, Hoboken, NJ, 2007},\
  \bibinfo {year} {2007})\ Chap.\ \bibinfo {chapter} {Variational Transition
  State Theory with Multidimensional Tunneling}, pp.\ \bibinfo {pages}
  {125--232},\ \bibinfo {edition} {1st}\ ed.\BibitemShut {Stop}%
\bibitem [{\citenamefont {Ochoa~de Aspuru}\ and\ \citenamefont
  {Clary}(1998)}]{och98}%
  \BibitemOpen
  \bibfield  {author} {\bibinfo {author} {\bibfnamefont {G.}~\bibnamefont
  {Ochoa~de Aspuru}}\ and\ \bibinfo {author} {\bibfnamefont {D.~C.}\
  \bibnamefont {Clary}},\ }\href {\doibase 10.1021/jp982433i} {\bibfield
  {journal} {\bibinfo  {journal} {J. Phys. Chem. A}\ }\textbf {\bibinfo
  {volume} {102}},\ \bibinfo {pages} {9631} (\bibinfo {year}
  {1998})}\BibitemShut {NoStop}%
\bibitem [{\citenamefont {Wu}\ \emph {et~al.}(2000)\citenamefont {Wu},
  \citenamefont {Schatz}, \citenamefont {Lendvay}, \citenamefont {Fang},\ and\
  \citenamefont {Harding}}]{wu00}%
  \BibitemOpen
  \bibfield  {author} {\bibinfo {author} {\bibfnamefont {G.-s.}\ \bibnamefont
  {Wu}}, \bibinfo {author} {\bibfnamefont {G.~C.}\ \bibnamefont {Schatz}},
  \bibinfo {author} {\bibfnamefont {G.}~\bibnamefont {Lendvay}}, \bibinfo
  {author} {\bibfnamefont {D.-C.}\ \bibnamefont {Fang}}, \ and\ \bibinfo
  {author} {\bibfnamefont {L.~B.}\ \bibnamefont {Harding}},\ }\href {\doibase
  http://dx.doi.org/10.1063/1.1287329} {\bibfield  {journal} {\bibinfo
  {journal} {J. Chem. Phys.}\ }\textbf {\bibinfo {volume} {113}},\ \bibinfo
  {pages} {3150} (\bibinfo {year} {2000})}\BibitemShut {NoStop}%
\bibitem [{\citenamefont {Walch}\ and\ \citenamefont {Dunning}(1980)}]{wal80}%
  \BibitemOpen
  \bibfield  {author} {\bibinfo {author} {\bibfnamefont {S.~P.}\ \bibnamefont
  {Walch}}\ and\ \bibinfo {author} {\bibfnamefont {T.~H.}\ \bibnamefont
  {Dunning}},\ }\href {\doibase http://dx.doi.org/10.1063/1.439193} {\bibfield
  {journal} {\bibinfo  {journal} {J. Chem. Phys.}\ }\textbf {\bibinfo {volume}
  {72}},\ \bibinfo {pages} {1303} (\bibinfo {year} {1980})}\BibitemShut
  {NoStop}%
\bibitem [{\citenamefont {Schatz}\ and\ \citenamefont
  {Elgersma}(1980)}]{sch80}%
  \BibitemOpen
  \bibfield  {author} {\bibinfo {author} {\bibfnamefont {G.~C.}\ \bibnamefont
  {Schatz}}\ and\ \bibinfo {author} {\bibfnamefont {H.}~\bibnamefont
  {Elgersma}},\ }\href {\doibase
  http://dx.doi.org/10.1016/0009-2614(80)85193-1} {\bibfield  {journal}
  {\bibinfo  {journal} {Chem. Phys. Lett.}\ }\textbf {\bibinfo {volume} {73}},\
  \bibinfo {pages} {21 } (\bibinfo {year} {1980})}\BibitemShut {NoStop}%
\bibitem [{\citenamefont {Yang}\ \emph {et~al.}(2001)\citenamefont {Yang},
  \citenamefont {Zhang}, \citenamefont {Collins},\ and\ \citenamefont
  {Lee}}]{yan01}%
  \BibitemOpen
  \bibfield  {author} {\bibinfo {author} {\bibfnamefont {M.}~\bibnamefont
  {Yang}}, \bibinfo {author} {\bibfnamefont {D.~H.}\ \bibnamefont {Zhang}},
  \bibinfo {author} {\bibfnamefont {M.~A.}\ \bibnamefont {Collins}}, \ and\
  \bibinfo {author} {\bibfnamefont {S.}~\bibnamefont {Lee}},\ }\href {\doibase
  10.1063/1.1372335} {\bibfield  {journal} {\bibinfo  {journal} {J. Chem.
  Phys.}\ }\textbf {\bibinfo {volume} {115}},\ \bibinfo {pages} {174} (\bibinfo
  {year} {2001})}\BibitemShut {NoStop}%
\bibitem [{\citenamefont {Bettens}\ \emph {et~al.}(2000)\citenamefont
  {Bettens}, \citenamefont {Collins}, \citenamefont {Jordan},\ and\
  \citenamefont {Zhang}}]{bet00}%
  \BibitemOpen
  \bibfield  {author} {\bibinfo {author} {\bibfnamefont {R.~P.~A.}\
  \bibnamefont {Bettens}}, \bibinfo {author} {\bibfnamefont {M.~A.}\
  \bibnamefont {Collins}}, \bibinfo {author} {\bibfnamefont {M.~J.~T.}\
  \bibnamefont {Jordan}}, \ and\ \bibinfo {author} {\bibfnamefont {D.~H.}\
  \bibnamefont {Zhang}},\ }\href {\doibase http://dx.doi.org/10.1063/1.481657}
  {\bibfield  {journal} {\bibinfo  {journal} {J. Chem. Phys.}\ }\textbf
  {\bibinfo {volume} {112}},\ \bibinfo {pages} {10162} (\bibinfo {year}
  {2000})}\BibitemShut {NoStop}%
\bibitem [{\citenamefont {Chen}\ \emph {et~al.}(2013)\citenamefont {Chen},
  \citenamefont {Xu}, \citenamefont {Xu},\ and\ \citenamefont {Zhang}}]{che13}%
  \BibitemOpen
  \bibfield  {author} {\bibinfo {author} {\bibfnamefont {J.}~\bibnamefont
  {Chen}}, \bibinfo {author} {\bibfnamefont {X.}~\bibnamefont {Xu}}, \bibinfo
  {author} {\bibfnamefont {X.}~\bibnamefont {Xu}}, \ and\ \bibinfo {author}
  {\bibfnamefont {D.~H.}\ \bibnamefont {Zhang}},\ }\href {\doibase
  http://dx.doi.org/10.1063/1.4801658} {\bibfield  {journal} {\bibinfo
  {journal} {J. Chem. Phys.}\ }\textbf {\bibinfo {volume} {138}},\ \bibinfo
  {eid} {154301} (\bibinfo {year} {2013})}\BibitemShut {NoStop}%
\bibitem [{\citenamefont {Matzkies}\ and\ \citenamefont
  {Manthe}(1998)}]{mat98}%
  \BibitemOpen
  \bibfield  {author} {\bibinfo {author} {\bibfnamefont {F.}~\bibnamefont
  {Matzkies}}\ and\ \bibinfo {author} {\bibfnamefont {U.}~\bibnamefont
  {Manthe}},\ }\href {\doibase http://dx.doi.org/10.1063/1.475892} {\bibfield
  {journal} {\bibinfo  {journal} {J. Chem. Phys.}\ }\textbf {\bibinfo {volume}
  {108}},\ \bibinfo {pages} {4828} (\bibinfo {year} {1998})}\BibitemShut
  {NoStop}%
\bibitem [{\citenamefont {Nguyen}, \citenamefont {Stanton},\ and\ \citenamefont
  {Barker}(2010)}]{ngu10}%
  \BibitemOpen
  \bibfield  {author} {\bibinfo {author} {\bibfnamefont {T.~L.}\ \bibnamefont
  {Nguyen}}, \bibinfo {author} {\bibfnamefont {J.~F.}\ \bibnamefont {Stanton}},
  \ and\ \bibinfo {author} {\bibfnamefont {J.~R.}\ \bibnamefont {Barker}},\
  }\href {\doibase http://dx.doi.org/10.1016/j.cplett.2010.09.015} {\bibfield
  {journal} {\bibinfo  {journal} {Chem. Phys. Lett.}\ }\textbf {\bibinfo
  {volume} {499}},\ \bibinfo {pages} {9 } (\bibinfo {year} {2010})}\BibitemShut
  {NoStop}%
\bibitem [{\citenamefont {Ree}, \citenamefont {Kim},\ and\ \citenamefont
  {Shin}(2015)}]{ree15}%
  \BibitemOpen
  \bibfield  {author} {\bibinfo {author} {\bibfnamefont {J.}~\bibnamefont
  {Ree}}, \bibinfo {author} {\bibfnamefont {Y.~H.}\ \bibnamefont {Kim}}, \ and\
  \bibinfo {author} {\bibfnamefont {H.~K.}\ \bibnamefont {Shin}},\ }\href@noop
  {} {\bibfield  {journal} {\bibinfo  {journal} {J. Phys. Chem. A}\ }\textbf
  {\bibinfo {volume} {119}},\ \bibinfo {pages} {3147} (\bibinfo {year}
  {2015})}\BibitemShut {NoStop}%
\bibitem [{\citenamefont {Brown}\ \emph {et~al.}(2001)\citenamefont {Brown},
  \citenamefont {Burkholder}, \citenamefont {Talukdar},\ and\ \citenamefont
  {Ravishankara}}]{bro01}%
  \BibitemOpen
  \bibfield  {author} {\bibinfo {author} {\bibfnamefont {S.~S.}\ \bibnamefont
  {Brown}}, \bibinfo {author} {\bibfnamefont {J.~B.}\ \bibnamefont
  {Burkholder}}, \bibinfo {author} {\bibfnamefont {R.~K.}\ \bibnamefont
  {Talukdar}}, \ and\ \bibinfo {author} {\bibfnamefont {A.~R.}\ \bibnamefont
  {Ravishankara}},\ }\href {\doibase 10.1021/jp002394m} {\bibfield  {journal}
  {\bibinfo  {journal} {J. Phys. Chem. A}\ }\textbf {\bibinfo {volume} {105}},\
  \bibinfo {pages} {1605} (\bibinfo {year} {2001})}\BibitemShut {NoStop}%
\bibitem [{\citenamefont {Shannon}\ \emph {et~al.}(2013)\citenamefont
  {Shannon}, \citenamefont {Blitz}, \citenamefont {Goddard},\ and\
  \citenamefont {Heard}}]{sha13}%
  \BibitemOpen
  \bibfield  {author} {\bibinfo {author} {\bibfnamefont {R.~J.}\ \bibnamefont
  {Shannon}}, \bibinfo {author} {\bibfnamefont {M.~A.}\ \bibnamefont {Blitz}},
  \bibinfo {author} {\bibfnamefont {A.}~\bibnamefont {Goddard}}, \ and\
  \bibinfo {author} {\bibfnamefont {D.~E.}\ \bibnamefont {Heard}},\ }\href
  {\doibase http://dx.doi.org/10.1038/nchem.1692} {\bibfield  {journal}
  {\bibinfo  {journal} {Nat. Chem.}\ }\textbf {\bibinfo {volume} {5}},\
  \bibinfo {pages} {745} (\bibinfo {year} {2013})}\BibitemShut {NoStop}%
\bibitem [{\citenamefont {Zhang}\ \emph {et~al.}(2014)\citenamefont {Zhang},
  \citenamefont {Rommel}, \citenamefont {Cvita{\v s}},\ and\ \citenamefont
  {Althorpe}}]{zha14}%
  \BibitemOpen
  \bibfield  {author} {\bibinfo {author} {\bibfnamefont {Y.}~\bibnamefont
  {Zhang}}, \bibinfo {author} {\bibfnamefont {J.~B.}\ \bibnamefont {Rommel}},
  \bibinfo {author} {\bibfnamefont {M.~T.}\ \bibnamefont {Cvita{\v s}}}, \ and\
  \bibinfo {author} {\bibfnamefont {S.~C.}\ \bibnamefont {Althorpe}},\ }\href
  {\doibase 10.1039/C4CP03235G} {\bibfield  {journal} {\bibinfo  {journal}
  {Phys. Chem. Chem. Phys.}\ }\textbf {\bibinfo {volume} {16}},\ \bibinfo
  {pages} {24292} (\bibinfo {year} {2014})}\BibitemShut {NoStop}%
\bibitem [{\citenamefont {Fern{\'a}ndez-Ramos}\ \emph
  {et~al.}(2007)\citenamefont {Fern{\'a}ndez-Ramos}, \citenamefont {Ellingson},
  \citenamefont {Meana-Pa{\~{n}}eda}, \citenamefont {Marques},\ and\
  \citenamefont {Truhlar}}]{fer07a}%
  \BibitemOpen
  \bibfield  {author} {\bibinfo {author} {\bibfnamefont {A.}~\bibnamefont
  {Fern{\'a}ndez-Ramos}}, \bibinfo {author} {\bibfnamefont {B.~A.}\
  \bibnamefont {Ellingson}}, \bibinfo {author} {\bibfnamefont {R.}~\bibnamefont
  {Meana-Pa{\~{n}}eda}}, \bibinfo {author} {\bibfnamefont {J.~M.~C.}\
  \bibnamefont {Marques}}, \ and\ \bibinfo {author} {\bibfnamefont {D.~G.}\
  \bibnamefont {Truhlar}},\ }\href {\doibase 10.1007/s00214-007-0328-0}
  {\bibfield  {journal} {\bibinfo  {journal} {Theor. Chem. Acc.}\ }\textbf
  {\bibinfo {volume} {118}},\ \bibinfo {pages} {813} (\bibinfo {year}
  {2007})}\BibitemShut {NoStop}%
\bibitem [{\citenamefont {K\"astner}\ \emph {et~al.}(2009)\citenamefont
  {K\"astner}, \citenamefont {Carr}, \citenamefont {Keal}, \citenamefont
  {Thiel}, \citenamefont {Wander},\ and\ \citenamefont {Sherwood}}]{kae09a}%
  \BibitemOpen
  \bibfield  {author} {\bibinfo {author} {\bibfnamefont {J.}~\bibnamefont
  {K\"astner}}, \bibinfo {author} {\bibfnamefont {J.~M.}\ \bibnamefont {Carr}},
  \bibinfo {author} {\bibfnamefont {T.~W.}\ \bibnamefont {Keal}}, \bibinfo
  {author} {\bibfnamefont {W.}~\bibnamefont {Thiel}}, \bibinfo {author}
  {\bibfnamefont {A.}~\bibnamefont {Wander}}, \ and\ \bibinfo {author}
  {\bibfnamefont {P.}~\bibnamefont {Sherwood}},\ }\href {\doibase
  10.1021/jp9028968} {\bibfield  {journal} {\bibinfo  {journal} {J. Phys. Chem.
  A}\ }\textbf {\bibinfo {volume} {113}},\ \bibinfo {pages} {11856} (\bibinfo
  {year} {2009})}\BibitemShut {NoStop}%
\bibitem [{\citenamefont {Adler}, \citenamefont {Knizia},\ and\ \citenamefont
  {Werner}(2007)}]{adl07}%
  \BibitemOpen
  \bibfield  {author} {\bibinfo {author} {\bibfnamefont {T.~B.}\ \bibnamefont
  {Adler}}, \bibinfo {author} {\bibfnamefont {G.}~\bibnamefont {Knizia}}, \
  and\ \bibinfo {author} {\bibfnamefont {H.-J.}\ \bibnamefont {Werner}},\
  }\href@noop {} {\bibfield  {journal} {\bibinfo  {journal} {J. Chem. Phys.}\
  }\textbf {\bibinfo {volume} {127}},\ \bibinfo {pages} {221106} (\bibinfo
  {year} {2007})}\BibitemShut {NoStop}%
\bibitem [{\citenamefont {Adler}\ and\ \citenamefont {Werner}(2009)}]{adl09}%
  \BibitemOpen
  \bibfield  {author} {\bibinfo {author} {\bibfnamefont {T.~B.}\ \bibnamefont
  {Adler}}\ and\ \bibinfo {author} {\bibfnamefont {H.-J.}\ \bibnamefont
  {Werner}},\ }\href {\doibase 10.1063/1.3160675} {\bibfield  {journal}
  {\bibinfo  {journal} {J. Chem. Phys.}\ }\textbf {\bibinfo {volume} {130}},\
  \bibinfo {eid} {241101} (\bibinfo {year} {2009})}\BibitemShut {NoStop}%
\bibitem [{\citenamefont {Peterson}, \citenamefont {Adler},\ and\ \citenamefont
  {Werner}(2008)}]{pet08}%
  \BibitemOpen
  \bibfield  {author} {\bibinfo {author} {\bibfnamefont {K.~A.}\ \bibnamefont
  {Peterson}}, \bibinfo {author} {\bibfnamefont {T.~B.}\ \bibnamefont {Adler}},
  \ and\ \bibinfo {author} {\bibfnamefont {H.-J.}\ \bibnamefont {Werner}},\
  }\href {\doibase http://dx.doi.org/10.1063/1.2831537} {\bibfield  {journal}
  {\bibinfo  {journal} {J. Chem. Phys.}\ }\textbf {\bibinfo {volume} {128}},\
  \bibinfo {eid} {084102} (\bibinfo {year} {2008})}\BibitemShut {NoStop}%
\bibitem [{\citenamefont {Werner}\ \emph
  {et~al.}(2012{\natexlab{a}})\citenamefont {Werner}, \citenamefont {Knowles},
  \citenamefont {Knizia}, \citenamefont {Manby}, \citenamefont {{Sch\"{u}tz}},
  \citenamefont {Celani}, \citenamefont {Korona}, \citenamefont {Lindh},
  \citenamefont {Mitrushenkov}, \citenamefont {Rauhut}, \citenamefont
  {Shamasundar}, \citenamefont {Adler}, \citenamefont {Amos}, \citenamefont
  {Bernhardsson}, \citenamefont {Berning}, \citenamefont {Cooper},
  \citenamefont {Deegan}, \citenamefont {Dobbyn}, \citenamefont {Eckert},
  \citenamefont {Goll}, \citenamefont {Hampel}, \citenamefont {Hesselmann},
  \citenamefont {Hetzer}, \citenamefont {Hrenar}, \citenamefont {Jansen},
  \citenamefont {K\"oppl}, \citenamefont {Liu}, \citenamefont {Lloyd},
  \citenamefont {Mata}, \citenamefont {May}, \citenamefont {McNicholas},
  \citenamefont {Meyer}, \citenamefont {Mura}, \citenamefont {Nicklass},
  \citenamefont {O'Neill}, \citenamefont {Palmieri}, \citenamefont {Peng},
  \citenamefont {Pfl\"uger}, \citenamefont {Pitzer}, \citenamefont {Reiher},
  \citenamefont {Shiozaki}, \citenamefont {Stoll}, \citenamefont {Stone},
  \citenamefont {Tarroni}, \citenamefont {Thorsteinsson},\ and\ \citenamefont
  {Wang}}]{MOLPRO2012}%
  \BibitemOpen
  \bibfield  {author} {\bibinfo {author} {\bibfnamefont {H.-J.}\ \bibnamefont
  {Werner}}, \bibinfo {author} {\bibfnamefont {P.~J.}\ \bibnamefont {Knowles}},
  \bibinfo {author} {\bibfnamefont {G.}~\bibnamefont {Knizia}}, \bibinfo
  {author} {\bibfnamefont {F.~R.}\ \bibnamefont {Manby}}, \bibinfo {author}
  {\bibfnamefont {M.}~\bibnamefont {{Sch\"{u}tz}}}, \bibinfo {author}
  {\bibfnamefont {P.}~\bibnamefont {Celani}}, \bibinfo {author} {\bibfnamefont
  {T.}~\bibnamefont {Korona}}, \bibinfo {author} {\bibfnamefont
  {R.}~\bibnamefont {Lindh}}, \bibinfo {author} {\bibfnamefont
  {A.}~\bibnamefont {Mitrushenkov}}, \bibinfo {author} {\bibfnamefont
  {G.}~\bibnamefont {Rauhut}}, \bibinfo {author} {\bibfnamefont {K.~R.}\
  \bibnamefont {Shamasundar}}, \bibinfo {author} {\bibfnamefont {T.~B.}\
  \bibnamefont {Adler}}, \bibinfo {author} {\bibfnamefont {R.~D.}\ \bibnamefont
  {Amos}}, \bibinfo {author} {\bibfnamefont {A.}~\bibnamefont {Bernhardsson}},
  \bibinfo {author} {\bibfnamefont {A.}~\bibnamefont {Berning}}, \bibinfo
  {author} {\bibfnamefont {D.~L.}\ \bibnamefont {Cooper}}, \bibinfo {author}
  {\bibfnamefont {M.~J.~O.}\ \bibnamefont {Deegan}}, \bibinfo {author}
  {\bibfnamefont {A.~J.}\ \bibnamefont {Dobbyn}}, \bibinfo {author}
  {\bibfnamefont {F.}~\bibnamefont {Eckert}}, \bibinfo {author} {\bibfnamefont
  {E.}~\bibnamefont {Goll}}, \bibinfo {author} {\bibfnamefont {C.}~\bibnamefont
  {Hampel}}, \bibinfo {author} {\bibfnamefont {A.}~\bibnamefont {Hesselmann}},
  \bibinfo {author} {\bibfnamefont {G.}~\bibnamefont {Hetzer}}, \bibinfo
  {author} {\bibfnamefont {T.}~\bibnamefont {Hrenar}}, \bibinfo {author}
  {\bibfnamefont {G.}~\bibnamefont {Jansen}}, \bibinfo {author} {\bibfnamefont
  {C.}~\bibnamefont {K\"oppl}}, \bibinfo {author} {\bibfnamefont
  {Y.}~\bibnamefont {Liu}}, \bibinfo {author} {\bibfnamefont {A.~W.}\
  \bibnamefont {Lloyd}}, \bibinfo {author} {\bibfnamefont {R.~A.}\ \bibnamefont
  {Mata}}, \bibinfo {author} {\bibfnamefont {A.~J.}\ \bibnamefont {May}},
  \bibinfo {author} {\bibfnamefont {S.~J.}\ \bibnamefont {McNicholas}},
  \bibinfo {author} {\bibfnamefont {W.}~\bibnamefont {Meyer}}, \bibinfo
  {author} {\bibfnamefont {M.~E.}\ \bibnamefont {Mura}}, \bibinfo {author}
  {\bibfnamefont {A.}~\bibnamefont {Nicklass}}, \bibinfo {author}
  {\bibfnamefont {D.~P.}\ \bibnamefont {O'Neill}}, \bibinfo {author}
  {\bibfnamefont {P.}~\bibnamefont {Palmieri}}, \bibinfo {author}
  {\bibfnamefont {D.}~\bibnamefont {Peng}}, \bibinfo {author} {\bibfnamefont
  {K.}~\bibnamefont {Pfl\"uger}}, \bibinfo {author} {\bibfnamefont
  {R.}~\bibnamefont {Pitzer}}, \bibinfo {author} {\bibfnamefont
  {M.}~\bibnamefont {Reiher}}, \bibinfo {author} {\bibfnamefont
  {T.}~\bibnamefont {Shiozaki}}, \bibinfo {author} {\bibfnamefont
  {H.}~\bibnamefont {Stoll}}, \bibinfo {author} {\bibfnamefont {A.~J.}\
  \bibnamefont {Stone}}, \bibinfo {author} {\bibfnamefont {R.}~\bibnamefont
  {Tarroni}}, \bibinfo {author} {\bibfnamefont {T.}~\bibnamefont
  {Thorsteinsson}}, \ and\ \bibinfo {author} {\bibfnamefont {M.}~\bibnamefont
  {Wang}},\ }\href@noop {} {\enquote {\bibinfo {title} {Molpro, version 2012.1,
  a package of ab initio programs},}\ } (\bibinfo {year}
  {2012}{\natexlab{a}})\BibitemShut {NoStop}%
\bibitem [{\citenamefont {Werner}\ \emph
  {et~al.}(2012{\natexlab{b}})\citenamefont {Werner}, \citenamefont {Knowles},
  \citenamefont {Knizia}, \citenamefont {Manby},\ and\ \citenamefont
  {Sch\"utz}}]{wer11}%
  \BibitemOpen
  \bibfield  {author} {\bibinfo {author} {\bibfnamefont {H.-J.}\ \bibnamefont
  {Werner}}, \bibinfo {author} {\bibfnamefont {P.~J.}\ \bibnamefont {Knowles}},
  \bibinfo {author} {\bibfnamefont {G.}~\bibnamefont {Knizia}}, \bibinfo
  {author} {\bibfnamefont {F.~R.}\ \bibnamefont {Manby}}, \ and\ \bibinfo
  {author} {\bibfnamefont {M.}~\bibnamefont {Sch\"utz}},\ }\href {\doibase
  10.1002/wcms.82} {\bibfield  {journal} {\bibinfo  {journal} {WIREs Comput.
  Mol. Sci.}\ }\textbf {\bibinfo {volume} {2}},\ \bibinfo {pages} {242}
  (\bibinfo {year} {2012}{\natexlab{b}})}\BibitemShut {NoStop}%
\bibitem [{\citenamefont {Metz}\ \emph {et~al.}(2014)\citenamefont {Metz},
  \citenamefont {K\"astner}, \citenamefont {Sokol}, \citenamefont {Keal},\ and\
  \citenamefont {Sherwood}}]{met14}%
  \BibitemOpen
  \bibfield  {author} {\bibinfo {author} {\bibfnamefont {S.}~\bibnamefont
  {Metz}}, \bibinfo {author} {\bibfnamefont {J.}~\bibnamefont {K\"astner}},
  \bibinfo {author} {\bibfnamefont {A.~A.}\ \bibnamefont {Sokol}}, \bibinfo
  {author} {\bibfnamefont {T.~W.}\ \bibnamefont {Keal}}, \ and\ \bibinfo
  {author} {\bibfnamefont {P.}~\bibnamefont {Sherwood}},\ }\href {\doibase
  10.1002/wcms.1163} {\bibfield  {journal} {\bibinfo  {journal} {WIREs Comput.
  Mol. Sci.}\ }\textbf {\bibinfo {volume} {4}},\ \bibinfo {pages} {101}
  (\bibinfo {year} {2014})}\BibitemShut {NoStop}%
\bibitem [{\citenamefont {Lu}\ \emph {et~al.}(1992)\citenamefont {Lu},
  \citenamefont {Truong}, \citenamefont {Melissas}, \citenamefont {Lynch},
  \citenamefont {Liu}, \citenamefont {Garrett}, \citenamefont {Steckler},
  \citenamefont {Isaacson}, \citenamefont {Rai}, \citenamefont {Hancock},
  \citenamefont {Lauderdale}, \citenamefont {Joseph},\ and\ \citenamefont
  {Truhlar}}]{lu92}%
  \BibitemOpen
  \bibfield  {author} {\bibinfo {author} {\bibfnamefont {D.}~\bibnamefont
  {Lu}}, \bibinfo {author} {\bibfnamefont {T.~N.}\ \bibnamefont {Truong}},
  \bibinfo {author} {\bibfnamefont {V.~S.}\ \bibnamefont {Melissas}}, \bibinfo
  {author} {\bibfnamefont {G.~C.}\ \bibnamefont {Lynch}}, \bibinfo {author}
  {\bibfnamefont {Y.}~\bibnamefont {Liu}}, \bibinfo {author} {\bibfnamefont
  {B.~C.}\ \bibnamefont {Garrett}}, \bibinfo {author} {\bibfnamefont
  {R.}~\bibnamefont {Steckler}}, \bibinfo {author} {\bibfnamefont {A.~D.}\
  \bibnamefont {Isaacson}}, \bibinfo {author} {\bibfnamefont {S.~N.}\
  \bibnamefont {Rai}}, \bibinfo {author} {\bibfnamefont {G.~C.}\ \bibnamefont
  {Hancock}}, \bibinfo {author} {\bibfnamefont {J.~G.}\ \bibnamefont
  {Lauderdale}}, \bibinfo {author} {\bibfnamefont {T.}~\bibnamefont {Joseph}},
  \ and\ \bibinfo {author} {\bibfnamefont {D.~G.}\ \bibnamefont {Truhlar}},\
  }\href {\doibase 10.1016/0010-4655(92)90012-N} {\bibfield  {journal}
  {\bibinfo  {journal} {Comput. Phys. Commun.}\ }\textbf {\bibinfo {volume}
  {71}},\ \bibinfo {pages} {235} (\bibinfo {year} {1992})}\BibitemShut
  {NoStop}%
\bibitem [{\citenamefont {Zheng}\ \emph {et~al.}(2010)\citenamefont {Zheng},
  \citenamefont {Zhang}, \citenamefont {Lynch}, \citenamefont {Corchado},
  \citenamefont {Chuang}, \citenamefont {Fast}, \citenamefont {Hu},
  \citenamefont {Liu}, \citenamefont {Lynch}, \citenamefont {Nguyen},
  \citenamefont {Jackels}, \citenamefont {Ramos}, \citenamefont {Ellingson},
  \citenamefont {Melissas}, \citenamefont {Vill\`{a}}, \citenamefont {Rossi},
  \citenamefont {{n}o}, \citenamefont {Pu}, \citenamefont {Albu}, \citenamefont
  {Steckler}, \citenamefont {Garrett}, \citenamefont {Isaacson},\ and\
  \citenamefont {Truhlar}}]{POLYRATE2010}%
  \BibitemOpen
  \bibfield  {author} {\bibinfo {author} {\bibfnamefont {J.}~\bibnamefont
  {Zheng}}, \bibinfo {author} {\bibfnamefont {S.}~\bibnamefont {Zhang}},
  \bibinfo {author} {\bibfnamefont {B.}~\bibnamefont {Lynch}}, \bibinfo
  {author} {\bibfnamefont {J.}~\bibnamefont {Corchado}}, \bibinfo {author}
  {\bibfnamefont {Y.-Y.}\ \bibnamefont {Chuang}}, \bibinfo {author}
  {\bibfnamefont {P.}~\bibnamefont {Fast}}, \bibinfo {author} {\bibfnamefont
  {W.-P.}\ \bibnamefont {Hu}}, \bibinfo {author} {\bibfnamefont {Y.-P.}\
  \bibnamefont {Liu}}, \bibinfo {author} {\bibfnamefont {G.}~\bibnamefont
  {Lynch}}, \bibinfo {author} {\bibfnamefont {K.}~\bibnamefont {Nguyen}},
  \bibinfo {author} {\bibfnamefont {C.}~\bibnamefont {Jackels}}, \bibinfo
  {author} {\bibfnamefont {A.~F.}\ \bibnamefont {Ramos}}, \bibinfo {author}
  {\bibfnamefont {B.}~\bibnamefont {Ellingson}}, \bibinfo {author}
  {\bibfnamefont {V.}~\bibnamefont {Melissas}}, \bibinfo {author}
  {\bibfnamefont {J.}~\bibnamefont {Vill\`{a}}}, \bibinfo {author}
  {\bibfnamefont {I.}~\bibnamefont {Rossi}}, \bibinfo {author} {\bibfnamefont
  {E.~C.}\ \bibnamefont {{n}o}}, \bibinfo {author} {\bibfnamefont
  {J.}~\bibnamefont {Pu}}, \bibinfo {author} {\bibfnamefont {T.}~\bibnamefont
  {Albu}}, \bibinfo {author} {\bibfnamefont {R.}~\bibnamefont {Steckler}},
  \bibinfo {author} {\bibfnamefont {B.}~\bibnamefont {Garrett}}, \bibinfo
  {author} {\bibfnamefont {A.}~\bibnamefont {Isaacson}}, \ and\ \bibinfo
  {author} {\bibfnamefont {D.}~\bibnamefont {Truhlar}},\ }\href@noop {}
  {\enquote {\bibinfo {title} {{POLYRATE-version 2010, University of Minnesota,
  Minneapolis, 2010.}}}\ } (\bibinfo {year} {2010})\BibitemShut {NoStop}%
\bibitem [{Thi()}]{ThisSI}%
  \BibitemOpen
  \href@noop {} {}\bibinfo {howpublished} {See supplemental material at [URL
  will be inserted by AIP] for tables with the values of the reaction rate
  constants and further comparison of CVT/$\mu$OMT and instanton reaction
  rates.}\BibitemShut {Stop}%
\bibitem [{\citenamefont {Defazio}\ and\ \citenamefont {Gray}(2003)}]{def03}%
  \BibitemOpen
  \bibfield  {author} {\bibinfo {author} {\bibfnamefont {P.}~\bibnamefont
  {Defazio}}\ and\ \bibinfo {author} {\bibfnamefont {S.~K.}\ \bibnamefont
  {Gray}},\ }\href {\doibase 10.1021/jp030190a} {\bibfield  {journal} {\bibinfo
   {journal} {J. Phys. Chem. A}\ }\textbf {\bibinfo {volume} {107}},\ \bibinfo
  {pages} {7132} (\bibinfo {year} {2003})}\BibitemShut {NoStop}%
\bibitem [{\citenamefont {{P\'{e}rez de Tudela}}\ \emph
  {et~al.}(2012)\citenamefont {{P\'{e}rez de Tudela}}, \citenamefont {Aoiz},
  \citenamefont {Suleimanov},\ and\ \citenamefont {Manolopoulos}}]{per12a}%
  \BibitemOpen
  \bibfield  {author} {\bibinfo {author} {\bibfnamefont {R.}~\bibnamefont
  {{P\'{e}rez de Tudela}}}, \bibinfo {author} {\bibfnamefont {F.~J.}\
  \bibnamefont {Aoiz}}, \bibinfo {author} {\bibfnamefont {Y.~V.}\ \bibnamefont
  {Suleimanov}}, \ and\ \bibinfo {author} {\bibfnamefont {D.~E.}\ \bibnamefont
  {Manolopoulos}},\ }\href {\doibase 10.1021/jz201702q} {\bibfield  {journal}
  {\bibinfo  {journal} {J. Phys. Chem. Lett.}\ }\textbf {\bibinfo {volume}
  {3}},\ \bibinfo {pages} {493} (\bibinfo {year} {2012})}\BibitemShut {NoStop}%
\bibitem [{\citenamefont {Suleimanov}\ \emph {et~al.}(2013)\citenamefont
  {Suleimanov}, \citenamefont {de~Tudela}, \citenamefont {Jambrina},
  \citenamefont {Castillo}, \citenamefont {S\'{a}ez-R\'{a}banos}, \citenamefont
  {Manolopoulos},\ and\ \citenamefont {Aoiz}}]{sul13}%
  \BibitemOpen
  \bibfield  {author} {\bibinfo {author} {\bibfnamefont {Y.~V.}\ \bibnamefont
  {Suleimanov}}, \bibinfo {author} {\bibfnamefont {R.~P.}\ \bibnamefont
  {de~Tudela}}, \bibinfo {author} {\bibfnamefont {P.~G.}\ \bibnamefont
  {Jambrina}}, \bibinfo {author} {\bibfnamefont {J.~F.}\ \bibnamefont
  {Castillo}}, \bibinfo {author} {\bibfnamefont {V.}~\bibnamefont
  {S\'{a}ez-R\'{a}banos}}, \bibinfo {author} {\bibfnamefont {D.~E.}\
  \bibnamefont {Manolopoulos}}, \ and\ \bibinfo {author} {\bibfnamefont
  {F.~J.}\ \bibnamefont {Aoiz}},\ }\href@noop {} {\bibfield  {journal}
  {\bibinfo  {journal} {Phys. Chem. Chem. Phys.}\ }\textbf {\bibinfo {volume}
  {15}},\ \bibinfo {pages} {3655} (\bibinfo {year} {2013})}\BibitemShut
  {NoStop}%
\bibitem [{\citenamefont {{P\'{e}rez de Tudela}}\ \emph
  {et~al.}(2014)\citenamefont {{P\'{e}rez de Tudela}}, \citenamefont
  {Suleimanov}, \citenamefont {Richardson}, \citenamefont {Rábanos},
  \citenamefont {Green},\ and\ \citenamefont {Aoiz}}]{per14}%
  \BibitemOpen
  \bibfield  {author} {\bibinfo {author} {\bibfnamefont {R.}~\bibnamefont
  {{P\'{e}rez de Tudela}}}, \bibinfo {author} {\bibfnamefont {Y.~V.}\
  \bibnamefont {Suleimanov}}, \bibinfo {author} {\bibfnamefont {J.~O.}\
  \bibnamefont {Richardson}}, \bibinfo {author} {\bibfnamefont {V.~S.}\
  \bibnamefont {Rábanos}}, \bibinfo {author} {\bibfnamefont {W.~H.}\
  \bibnamefont {Green}}, \ and\ \bibinfo {author} {\bibfnamefont {F.~J.}\
  \bibnamefont {Aoiz}},\ }\href {\doibase 10.1021/jz502216g} {\bibfield
  {journal} {\bibinfo  {journal} {J. Phys. Chem. Lett.}\ }\textbf {\bibinfo
  {volume} {5}},\ \bibinfo {pages} {4219} (\bibinfo {year} {2014})}\BibitemShut
  {NoStop}%
\end{thebibliography}

%


\clearpage
\bf{
\noindent
Supplementary Information to: 
Reaction Rates and Kinetic Isotope Effects of
H$_2$ + OH $\rightarrow$ H$_2$O + H
}
\vfill


%
%



\maketitle

\begin{figure}[h!]
  \begin{center}
    \includegraphics[width=16cm]{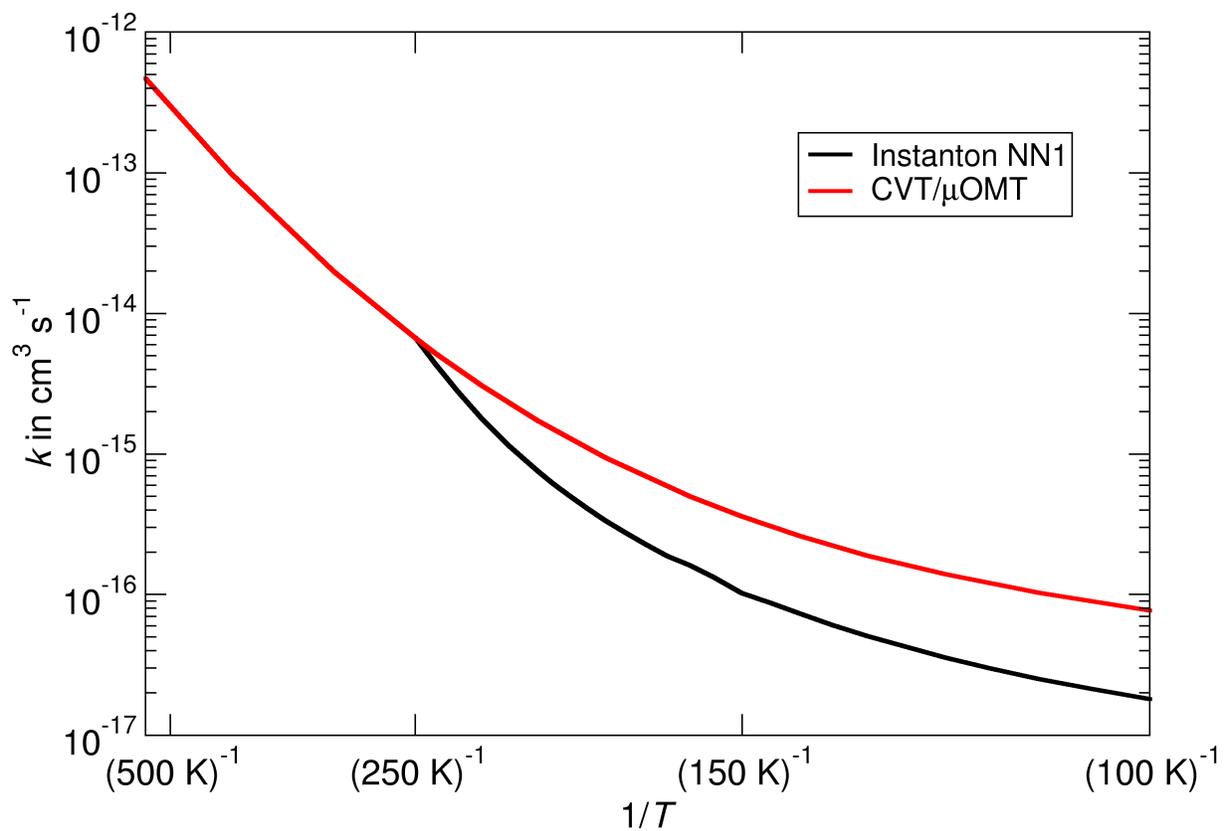}
    \caption{Comparison of CVT/$\mu$OMT and instanton reaction rate constants over the 
whole range of applicability of instanton theory. 
      \label{fig:comparison}
    }
  \end{center}
\end{figure}

\begin{table}[h!]
 \caption{
 \label{allrates}
Reaction rate constants for the H-transfer calculated with the instanton method on the NN1 PES.
Temperature in K, rate constants in cm$^3$ molecule$^{-1}$ s$^{-1}$.
  }								
  \begin{center}
\setlength{\tabcolsep}{4mm}
\begin{tabular}{rrrrrrrrrrr}
\hline
Temperature     &       HHOH    &       DHOH    &       HHOD    &       DHOD    \\\hline
100     &$      1.81 \cdot 10^{ -17     }$&$    4.58 \cdot 10^{ -18     }$&$    3.22 \cdot 10^{ -17     }$&$    8.86 \cdot 10^{ -18     }$\\
105     &$      2.13 \cdot 10^{ -17     }$&$    5.49 \cdot 10^{ -18     }$&$    3.73 \cdot 10^{ -17     }$&$    1.04 \cdot 10^{ -17     }$\\
110     &$      2.51 \cdot 10^{ -17     }$&$    6.61 \cdot 10^{ -18     }$&$    4.34 \cdot 10^{ -17     }$&$    1.23 \cdot 10^{ -17     }$\\
115     &$      2.98 \cdot 10^{ -17     }$&$    8.01 \cdot 10^{ -18     }$&$    5.08 \cdot 10^{ -17     }$&$    1.47 \cdot 10^{ -17     }$\\
120     &$      3.55 \cdot 10^{ -17     }$&$    9.73 \cdot 10^{ -18     }$&$    5.96 \cdot 10^{ -17     }$&$    1.75 \cdot 10^{ -17     }$\\
130     &$      5.06 \cdot 10^{ -17     }$&$    1.45 \cdot 10^{ -17     }$&$    8.30 \cdot 10^{ -17     }$&$    2.53 \cdot 10^{ -17     }$\\
135     &$      6.07 \cdot 10^{ -17     }$&$    1.78 \cdot 10^{ -17     }$&$    9.84 \cdot 10^{ -17     }$&$    3.06 \cdot 10^{ -17     }$\\
140     &$      7.30 \cdot 10^{ -17     }$&$    2.19 \cdot 10^{ -17     }$&$    1.17 \cdot 10^{ -16     }$&$    3.71 \cdot 10^{ -17     }$\\
145     &$      8.77 \cdot 10^{ -17     }$&$    2.73 \cdot 10^{ -17     }$&$    1.39 \cdot 10^{ -16     }$&$    4.55 \cdot 10^{ -17     }$\\
150     &$      1.02 \cdot 10^{ -16     }$&$    3.54 \cdot 10^{ -17     }$&$    1.66 \cdot 10^{ -16     }$&$    5.83 \cdot 10^{ -17     }$\\
155     &$      1.31 \cdot 10^{ -16     }$&$    4.15 \cdot 10^{ -17     }$&$    2.04 \cdot 10^{ -16     }$&$    6.74 \cdot 10^{ -17     }$\\
160     &$      1.61 \cdot 10^{ -16     }$&$    5.06 \cdot 10^{ -17     }$&$    2.49 \cdot 10^{ -16     }$&$    8.12 \cdot 10^{ -17     }$\\
165     &$      1.88 \cdot 10^{ -16     }$&$    6.30 \cdot 10^{ -17     }$&$    2.87 \cdot 10^{ -16     }$&$    9.99 \cdot 10^{ -17     }$\\
170     &$      2.27 \cdot 10^{ -16     }$&$    7.86 \cdot 10^{ -17     }$&$    3.43 \cdot 10^{ -16     }$&$    1.23 \cdot 10^{ -16     }$\\
175     &$      2.76 \cdot 10^{ -16     }$&$    9.80 \cdot 10^{ -17     }$&$    4.14 \cdot 10^{ -16     }$&$    1.52 \cdot 10^{ -16     }$\\
180     &$      3.36 \cdot 10^{ -16     }$&$    1.22 \cdot 10^{ -16     }$&$    5.00 \cdot 10^{ -16     }$&$    1.87 \cdot 10^{ -16     }$\\
185     &$      4.10 \cdot 10^{ -16     }$&$    1.52 \cdot 10^{ -16     }$&$    6.06 \cdot 10^{ -16     }$&$    2.31 \cdot 10^{ -16     }$\\
190     &$      5.02 \cdot 10^{ -16     }$&$    1.90 \cdot 10^{ -16     }$&$    7.36 \cdot 10^{ -16     }$&$    2.86 \cdot 10^{ -16     }$\\
195     &$      6.15 \cdot 10^{ -16     }$&$    2.38 \cdot 10^{ -16     }$&$    8.96 \cdot 10^{ -16     }$&$    3.54 \cdot 10^{ -16     }$\\
200     &$      7.56 \cdot 10^{ -16     }$&$    2.98 \cdot 10^{ -16     }$&$    1.09 \cdot 10^{ -15     }$&$    4.40 \cdot 10^{ -16     }$\\
210     &$      1.15 \cdot 10^{ -15     }$&$    4.73 \cdot 10^{ -16     }$&$    1.64 \cdot 10^{ -15     }$&$    6.88 \cdot 10^{ -16     }$\\
220     &$      1.78 \cdot 10^{ -15     }$&$    7.63 \cdot 10^{ -16     }$&$    2.51 \cdot 10^{ -15     }$&$    1.10 \cdot 10^{ -15     }$\\
230     &$      2.78 \cdot 10^{ -15     }$&$    1.24 \cdot 10^{ -15     }$&$    3.90 \cdot 10^{ -15     }$&$    1.77 \cdot 10^{ -15     }$\\
240     &$      4.37 \cdot 10^{ -15     }$&$    2.00 \cdot 10^{ -15     }$&$    6.07 \cdot 10^{ -15     }$&$    2.82 \cdot 10^{ -15     }$\\
250     &$      6.71 \cdot 10^{ -15     }$&$    3.10 \cdot 10^{ -15     }$&$    9.26 \cdot 10^{ -15     }$&$    4.34 \cdot 10^{ -15     }$\\
\hline
\end{tabular}
\end{center}
\end{table}

\begin{table}[h!]
 \caption{
 \label{allrates_d}
Reaction rate constants for the D-transfer calculated with the instanton method on the NN1 PES.
Temperature in K, rate constants in cm$^3$ molecule$^{-1}$ s$^{-1}$.
  }								
  \begin{center}
\setlength{\tabcolsep}{4mm}
\begin{tabular}{rrrrrrrrrrr}
\hline
$T$ [K]     &       HDOH    &       DDOH    &       HDOD    &       DDOD    \\\hline
80      &$      2.70 \cdot 10^{ -20     }$&$    2.50 \cdot 10^{ -20     }$&$    6.14 \cdot 10^{ -20     }$&$    6.02 \cdot 10^{ -20     }$\\
84      &$      3.32 \cdot 10^{ -20     }$&$    3.11 \cdot 10^{ -20     }$&$    7.39 \cdot 10^{ -20     }$&$    7.31 \cdot 10^{ -20     }$\\
88      &$      4.11 \cdot 10^{ -20     }$&$    3.90 \cdot 10^{ -20     }$&$    8.96 \cdot 10^{ -20     }$&$    8.97 \cdot 10^{ -20     }$\\
92      &$      5.11 \cdot 10^{ -20     }$&$    4.93 \cdot 10^{ -20     }$&$    1.10 \cdot 10^{ -19     }$&$    1.11 \cdot 10^{ -19     }$\\
96      &$      6.40 \cdot 10^{ -20     }$&$    6.27 \cdot 10^{ -20     }$&$    1.35 \cdot 10^{ -19     }$&$    1.39 \cdot 10^{ -19     }$\\
100     &$      8.05 \cdot 10^{ -20     }$&$    8.03 \cdot 10^{ -20     }$&$    1.67 \cdot 10^{ -19     }$&$    1.74 \cdot 10^{ -19     }$\\
105     &$      1.08 \cdot 10^{ -19     }$&$    1.10 \cdot 10^{ -19     }$&$    2.19 \cdot 10^{ -19     }$&$    2.33 \cdot 10^{ -19     }$\\
110     &$      1.45 \cdot 10^{ -19     }$&$    1.51 \cdot 10^{ -19     }$&$    2.89 \cdot 10^{ -19     }$&$    3.15 \cdot 10^{ -19     }$\\
115     &$      1.96 \cdot 10^{ -19     }$&$    2.19 \cdot 10^{ -19     }$&$    3.83 \cdot 10^{ -19     }$&$    4.47 \cdot 10^{ -19     }$\\
120     &$      2.80 \cdot 10^{ -19     }$&$    3.00 \cdot 10^{ -19     }$&$    5.41 \cdot 10^{ -19     }$&$    6.01 \cdot 10^{ -19     }$\\
130     &$      5.05 \cdot 10^{ -19     }$&$    5.89 \cdot 10^{ -19     }$&$    9.47 \cdot 10^{ -19     }$&$    1.14 \cdot 10^{ -18     }$\\
135     &$      6.96 \cdot 10^{ -19     }$&$    8.39 \cdot 10^{ -19     }$&$    1.29 \cdot 10^{ -18     }$&$    1.73 \cdot 10^{ -18     }$\\
140     &$      9.69 \cdot 10^{ -19     }$&$    1.20 \cdot 10^{ -18     }$&$    1.77 \cdot 10^{ -18     }$&$    2.25 \cdot 10^{ -18     }$\\
145     &$      1.36 \cdot 10^{ -18     }$&$    1.73 \cdot 10^{ -18     }$&$    2.44 \cdot 10^{ -18     }$&$    3.19 \cdot 10^{ -18     }$\\
150     &$      1.91 \cdot 10^{ -18     }$&$    2.50 \cdot 10^{ -18     }$&$    3.40 \cdot 10^{ -18     }$&$    4.55 \cdot 10^{ -18     }$\\
155     &$      2.72 \cdot 10^{ -18     }$&$    4.00 \cdot 10^{ -18     }$&$    4.77 \cdot 10^{ -18     }$&$    6.53 \cdot 10^{ -18     }$\\
160     &$      3.88 \cdot 10^{ -18     }$&$    5.33 \cdot 10^{ -18     }$&$    6.74 \cdot 10^{ -18     }$&$    9.44 \cdot 10^{ -18     }$\\
165     &$      5.59 \cdot 10^{ -18     }$&$    7.84 \cdot 10^{ -18     }$&$    9.61 \cdot 10^{ -18     }$&$    1.37 \cdot 10^{ -17     }$\\
170     &$      8.08 \cdot 10^{ -18     }$&$    1.16 \cdot 10^{ -17     }$&$    1.38 \cdot 10^{ -17     }$&$    2.01 \cdot 10^{ -17     }$\\
175     &$      1.17 \cdot 10^{ -17     }$&$    1.71 \cdot 10^{ -17     }$&$    1.99 \cdot 10^{ -17     }$&$    2.94 \cdot 10^{ -17     }$\\
180     &$      1.71 \cdot 10^{ -17     }$&$    2.51 \cdot 10^{ -17     }$&$    2.86 \cdot 10^{ -17     }$&$    4.28 \cdot 10^{ -17     }$\\
185     &$      2.46 \cdot 10^{ -17     }$&$    3.64 \cdot 10^{ -17     }$&$    4.10 \cdot 10^{ -17     }$&$    6.16 \cdot 10^{ -17     }$\\
190     &$      3.50 \cdot 10^{ -17     }$&$    5.22 \cdot 10^{ -17     }$&$    5.80 \cdot 10^{ -17     }$&$    8.78 \cdot 10^{ -17     }$\\
195     &$      4.93 \cdot 10^{ -17     }$&$    7.46 \cdot 10^{ -17     }$&$    8.11 \cdot 10^{ -17     }$&$    1.25 \cdot 10^{ -16     }$\\
\hline
\end{tabular}
\end{center}
\end{table}

\begin{table}[h!]
 \caption{
 \label{ontheflyrates}
Reaction rate constants for the HHOH system calculated on-the-fly on CCSD(T)-F12/cc-pVDZ-F12 level
with the instanton method.
Temperature in K, rate constants in cm$^3$ molecule$^{-1}$ s$^{-1}$.
  }								
  \begin{center}
\setlength{\tabcolsep}{4mm}
\begin{tabular}{rrrrr}
\hline
$T$ [K]& Rate constants \\
\hline
$       275     $&$     1.38\cdot 10^{ -14      }$\\
$       270     $&$     9.38\cdot 10^{ -15      }$\\
$       260     $&$     4.50\cdot 10^{ -15      }$\\
$       240     $&$     2.17\cdot 10^{ -15      }$\\
$       230     $&$     1.52\cdot 10^{ -15      }$\\
$       220     $&$     9.74\cdot 10^{ -16      }$\\
$       210     $&$     5.88\cdot 10^{ -16      }$\\
$       200     $&$     4.83\cdot 10^{ -16      }$\\
$       180     $&$     1.66\cdot 10^{ -16      }$\\
$       170     $&$     1.34\cdot 10^{ -16      }$\\
$       162     $&$     9.00\cdot 10^{ -17      }$\\
$       145     $&$     3.89\cdot 10^{ -17      }$\\
$       125     $&$     1.76\cdot 10^{ -17      }$\\
$       111     $&$     1.09\cdot 10^{ -17      }$\\
$       100     $&$     7.80\cdot 10^{ -18      }$\\
\hline
\end{tabular}
\end{center}
\end{table}

\begin{table}[h!]
 \caption{
 \label{muomtrates_h}
Reaction rate constants for the H-transfer calculated with CVT/$\mu$OMT on the NN1 PES.
Temperature in K, rate constants in cm$^3$ molecule$^{-1}$ s$^{-1}$.
  }
  \begin{center}
\setlength{\tabcolsep}{4mm}
\begin{tabular}{rrrrrrrrrrrrr}
\hline
   $T$ [K]        &                       HHOH    &                      HHOD    &                   DHOH    &       DHOD                     \\
\hline
$	    50.00 $ &$  2.50\cdot 10^{-17	}$	&	$	    4.10\cdot 10^{-17 	}$	&	$	     3.63\cdot 10^{-18  	}$	&	$	      7.30\cdot 10^{-18  	}$		\\
$	    55.00 $ &$  2.69\cdot 10^{-17	}$	&	$	    4.32\cdot 10^{-17 	}$	&	$	     4.02\cdot 10^{-18  	}$	&	$	      7.85\cdot 10^{-18  	}$		\\
$	    60.00 $ &$  2.94\cdot 10^{-17	}$	&	$	    4.61\cdot 10^{-17 	}$	&	$	     4.51\cdot 10^{-18  	}$	&	$	      8.56\cdot 10^{-18  	}$		\\
$	    65.00 $ &$  3.24\cdot 10^{-17	}$	&	$	    4.97\cdot 10^{-17 	}$	&	$	     5.12\cdot 10^{-18  	}$	&	$	      9.44\cdot 10^{-18  	}$		\\
$	    70.00 $ &$  3.60\cdot 10^{-17	}$	&	$	    5.42\cdot 10^{-17 	}$	&	$	     5.87\cdot 10^{-18  	}$	&	$	      1.05\cdot 10^{-17  	}$		\\
$	    75.00 $ &$  4.03\cdot 10^{-17	}$	&	$	    5.95\cdot 10^{-17 	}$	&	$	     6.78\cdot 10^{-18  	}$	&	$	      1.19\cdot 10^{-17  	}$		\\
$	    80.00 $ &$  4.54\cdot 10^{-17	}$	&	$	    6.58\cdot 10^{-17 	}$	&	$	     7.89\cdot 10^{-18  	}$	&	$	      1.35\cdot 10^{-17  	}$		\\
$	    90.00 $ &$  5.85\cdot 10^{-17	}$	&	$	    8.23\cdot 10^{-17 	}$	&	$	     1.09\cdot 10^{-17  	}$	&	$	      1.78\cdot 10^{-17  	}$		\\
$	   100.00 $ &$  7.69\cdot 10^{-17	}$	&	$	    1.05\cdot 10^{-16 	}$	&	$	     1.54\cdot 10^{-17  	}$	&	$	      2.41\cdot 10^{-17  	}$		\\
$	   110.00 $ &$  1.03\cdot 10^{-16	}$	&	$	    1.37\cdot 10^{-16 	}$	&	$	     2.21\cdot 10^{-17  	}$	&	$	      3.34\cdot 10^{-17  	}$		\\
$	   120.00 $ &$  1.39\cdot 10^{-16	}$	&	$	    1.81\cdot 10^{-16 	}$	&	$	     3.21\cdot 10^{-17  	}$	&	$	      4.68\cdot 10^{-17  	}$		\\
$	   130.00 $ &$  1.89\cdot 10^{-16	}$	&	$	    2.42\cdot 10^{-16 	}$	&	$	     4.69\cdot 10^{-17  	}$	&	$	      6.63\cdot 10^{-17  	}$		\\
$	   140.00 $ &$  2.60\cdot 10^{-16	}$	&	$	    3.26\cdot 10^{-16 	}$	&	$	     6.89\cdot 10^{-17  	}$	&	$	      9.46\cdot 10^{-17  	}$		\\
$	   150.00 $ &$  3.59\cdot 10^{-16	}$	&	$	    4.42\cdot 10^{-16 	}$	&	$	     1.01\cdot 10^{-16  	}$	&	$	      1.35\cdot 10^{-16  	}$		\\
$	   160.00 $ &$  4.96\cdot 10^{-16	}$	&	$	    5.99\cdot 10^{-16 	}$	&	$	     1.47\cdot 10^{-16  	}$	&	$	      1.92\cdot 10^{-16  	}$		\\
$	   180.00 $ &$  9.38\cdot 10^{-16	}$	&	$	    1.09\cdot 10^{-15 	}$	&	$	     3.06\cdot 10^{-16  	}$	&	$	      3.83\cdot 10^{-16  	}$		\\
$	   200.00 $ &$  1.73\cdot 10^{-15	}$	&	$	    1.95\cdot 10^{-15 	}$	&	$	     6.05\cdot 10^{-16  	}$	&	$	      7.33\cdot 10^{-16  	}$		\\
$	   220.00 $ &$  3.06\cdot 10^{-15	}$	&	$	    3.36\cdot 10^{-15 	}$	&	$	     1.14\cdot 10^{-15  	}$	&	$	      1.33\cdot 10^{-15  	}$		\\
$	   240.00 $ &$  5.20\cdot 10^{-15	}$	&	$	    5.56\cdot 10^{-15 	}$	&	$	     2.01\cdot 10^{-15  	}$	&	$	      2.28\cdot 10^{-15  	}$		\\
$	   250.00 $ &$  6.68\cdot 10^{-15	}$	&	$	    7.06\cdot 10^{-15 	}$	&	$	     2.61\cdot 10^{-15  	}$	&	$	      2.94\cdot 10^{-15  	}$		\\
$	   298.00 $ &$  1.92\cdot 10^{-14	}$	&	$	    1.94\cdot 10^{-14 	}$	&	$	     7.94\cdot 10^{-15  	}$	&	$	      8.45\cdot 10^{-15  	}$		\\
$	   300.00 $ &$  2.00\cdot 10^{-14	}$	&	$	    2.01\cdot 10^{-14 	}$	&	$	     8.20\cdot 10^{-15  	}$	&	$	      8.79\cdot 10^{-15  	}$		\\
$	   400.00 $ &$  9.87\cdot 10^{-14	}$	&	$	    9.33\cdot 10^{-14 	}$	&	$	     4.17\cdot 10^{-14  	}$	&	$	      4.17\cdot 10^{-14  	}$		\\
$	   600.00 $ &$  6.30\cdot 10^{-13	}$	&	$	    5.86\cdot 10^{-13 	}$	&	$	     2.76\cdot 10^{-13  	}$	&	$	      2.60\cdot 10^{-13  	}$		\\
$	   800.00 $ &$  2.02\cdot 10^{-12	}$	&	$	    1.76\cdot 10^{-12 	}$	&	$	     8.48\cdot 10^{-13  	}$	&	$	      7.77\cdot 10^{-13  	}$		\\
$	  1000.00 $ &$  4.35\cdot 10^{-12	}$	&	$	    3.76\cdot 10^{-12 	}$	&	$	     1.86\cdot 10^{-12  	}$	&	$	      1.67\cdot 10^{-12  	}$		\\
$	  1250.00 $ &$  8.96\cdot 10^{-12	}$	&	$	    7.61\cdot 10^{-12 	}$	&	$	     3.90\cdot 10^{-12  	}$	&	$	      3.49\cdot 10^{-12  	}$		\\
$	  1500.00 $ &$  1.56\cdot 10^{-11	}$	&	$	    1.32\cdot 10^{-11 	}$	&	$	     6.92\cdot 10^{-12  	}$	&	$	      6.20\cdot 10^{-12  	}$		\\
$	  2000.00 $ &$  3.63\cdot 10^{-11	}$	&	$	    3.06\cdot 10^{-11 	}$	&	$	     1.64\cdot 10^{-11  	}$	&	$	      1.47\cdot 10^{-11  	}$		\\
\hline
\end{tabular}
\end{center}
\end{table}

\begin{table}[h!]
 \caption{
 \label{muomtrates_h}
Reaction rate constants for the D-transfer calculated with CVT/$\mu$OMT on the NN1 PES.
Temperature in K, rate constants in cm$^3$ molecule$^{-1}$ s$^{-1}$.
  }
  \begin{center}
\setlength{\tabcolsep}{4mm}
\begin{tabular}{rrrrrrrrrrr}
\hline
   $T$ [K]        &                       HDOH    &                      HDOD    &                   DDOH    &       DDOD                     \\
\hline
$	    50.00 $&$6.54 \cdot 10^{-20}$	&	$	     1.16\cdot 10^{-19  	}$	&	$	    4.48\cdot 10^{-20  	}$	&	$  1.09\cdot 10^{-19  	}$	\\	     
$	    55.00 $&$7.42 \cdot 10^{-20}$	&	$	     1.30\cdot 10^{-19  	}$	&	$	    5.25\cdot 10^{-20  	}$	&	$  1.23\cdot 10^{-19  	}$	\\	     
$	    60.00 $&$8.55 \cdot 10^{-20}$	&	$	     1.49\cdot 10^{-19  	}$	&	$	    6.28\cdot 10^{-20  	}$	&	$  1.43\cdot 10^{-19  	}$	\\	     
$	    65.00 $&$1.00 \cdot 10^{-19}$	&	$	     1.74\cdot 10^{-19  	}$	&	$	    7.65\cdot 10^{-20  	}$	&	$  1.69\cdot 10^{-19  	}$	\\	     
$	    70.00 $&$1.19 \cdot 10^{-19}$	&	$	     2.05\cdot 10^{-19  	}$	&	$	    9.49\cdot 10^{-20  	}$	&	$  2.02\cdot 10^{-19  	}$	\\	     
$	    75.00 $&$1.43 \cdot 10^{-19}$	&	$	     2.46\cdot 10^{-19  	}$	&	$	    1.19\cdot 10^{-19  	}$	&	$  2.47\cdot 10^{-19  	}$	\\	     
$	    80.00 $&$1.75 \cdot 10^{-19}$	&	$	     2.98\cdot 10^{-19  	}$	&	$	    1.52\cdot 10^{-19  	}$	&	$  3.06\cdot 10^{-19  	}$	\\	     
$	    90.00 $&$2.72 \cdot 10^{-19}$	&	$	     4.52\cdot 10^{-19  	}$	&	$	    2.56\cdot 10^{-19  	}$	&	$  4.87\cdot 10^{-19  	}$	\\	     
$	   100.00 $&$4.38 \cdot 10^{-19}$	&	$	     7.08\cdot 10^{-19  	}$	&	$	    4.46\cdot 10^{-19  	}$	&	$  8.07\cdot 10^{-19  	}$	\\	     
$	   110.00 $&$7.25 \cdot 10^{-19}$	&	$	     1.14\cdot 10^{-18  	}$	&	$	    7.99\cdot 10^{-19  	}$	&	$  1.38\cdot 10^{-18  	}$	\\	     
$	   120.00 $&$1.22 \cdot 10^{-18}$	&	$	     1.87\cdot 10^{-18  	}$	&	$	    1.45\cdot 10^{-18  	}$	&	$  2.41\cdot 10^{-18  	}$	\\	     
$	   130.00 $&$2.08 \cdot 10^{-18}$	&	$	     3.09\cdot 10^{-18  	}$	&	$	    2.66\cdot 10^{-18  	}$	&	$  4.24\cdot 10^{-18  	}$	\\	     
$	   140.00 $&$3.55 \cdot 10^{-18}$	&	$	     5.14\cdot 10^{-18  	}$	&	$	    4.83\cdot 10^{-18  	}$	&	$  7.44\cdot 10^{-18  	}$	\\	     
$	   150.00 $&$6.03 \cdot 10^{-18}$	&	$	     8.52\cdot 10^{-18  	}$	&	$	    8.68\cdot 10^{-18  	}$	&	$  1.30\cdot 10^{-17  	}$	\\	     
$	   160.00 $&$1.01 \cdot 10^{-17}$	&	$	     1.40\cdot 10^{-17  	}$	&	$	    1.53\cdot 10^{-17  	}$	&	$  2.22\cdot 10^{-17  	}$	\\	     
$	   180.00 $&$2.75 \cdot 10^{-17}$	&	$	     3.61\cdot 10^{-17  	}$	&	$	    4.44\cdot 10^{-17  	}$	&	$  6.08\cdot 10^{-17  	}$	\\	     
$	   200.00 $&$6.93 \cdot 10^{-17}$	&	$	     8.69\cdot 10^{-17  	}$	&	$	    1.16\cdot 10^{-16  	}$	&	$  1.52\cdot 10^{-16  	}$	\\	     
$	   220.00 $&$1.60 \cdot 10^{-16}$	&	$	     1.92\cdot 10^{-16  	}$	&	$	    2.72\cdot 10^{-16  	}$	&	$  3.41\cdot 10^{-16  	}$	\\	     
$	   240.00 $&$3.40 \cdot 10^{-16}$	&	$	     3.94\cdot 10^{-16  	}$	&	$	    5.81\cdot 10^{-16  	}$	&	$  7.05\cdot 10^{-16  	}$	\\	     
$	   250.00 $&$4.82 \cdot 10^{-16}$	&	$	     5.49\cdot 10^{-16  	}$	&	$	    8.24\cdot 10^{-16  	}$	&	$  9.85\cdot 10^{-16  	}$	\\	     
$	   298.00 $&$2.02 \cdot 10^{-15}$	&	$	     2.14\cdot 10^{-15  	}$	&	$	    3.41\cdot 10^{-15  	}$	&	$  3.81\cdot 10^{-15  	}$	\\	     
$	   300.00 $&$2.13 \cdot 10^{-15}$	&	$	     2.25\cdot 10^{-15  	}$	&	$	    3.60\cdot 10^{-15  	}$	&	$  4.01\cdot 10^{-15  	}$	\\	     
$	   400.00 $&$1.68 \cdot 10^{-14}$	&	$	     1.63\cdot 10^{-14  	}$	&	$	    2.72\cdot 10^{-14  	}$	&	$  2.74\cdot 10^{-14  	}$	\\	     
$	   600.00 $&$1.72 \cdot 10^{-13}$	&	$	     1.55\cdot 10^{-13  	}$	&	$	    2.70\cdot 10^{-13  	}$	&	$  2.57\cdot 10^{-13  	}$	\\	     
$	   800.00 $&$6.42 \cdot 10^{-13}$	&	$	     5.59\cdot 10^{-13  	}$	&	$	    1.01\cdot 10^{-12  	}$	&	$  9.36\cdot 10^{-13  	}$	\\	     
$	  1000.00 $&$1.56 \cdot 10^{-12}$	&	$	     1.34\cdot 10^{-12  	}$	&	$	    2.50\cdot 10^{-12  	}$	&	$  2.28\cdot 10^{-12  	}$	\\	     
$	  1250.00 $&$3.50 \cdot 10^{-12}$	&	$	     2.98\cdot 10^{-12  	}$	&	$	    5.71\cdot 10^{-12  	}$	&	$  5.16\cdot 10^{-12  	}$	\\	     
$	  1500.00 $&$6.45 \cdot 10^{-12}$	&	$	     5.44\cdot 10^{-12  	}$	&	$	    1.08\cdot 10^{-11  	}$	&	$  9.61\cdot 10^{-12  	}$	\\	     
$	  2000.00 $&$1.60 \cdot 10^{-11}$	&	$	     1.34\cdot 10^{-11  	}$	&	$	    2.71\cdot 10^{-11  	}$	&	$  2.40\cdot 10^{-11  	}$	\\	     
\hline
\end{tabular}
\end{center}
\end{table}

\end{document}